\documentclass[a4paper,fleqn,usenatbib]{mnras}
\usepackage[T1]{fontenc}
\usepackage{ae,aecompl}
\usepackage{graphicx}	
\usepackage{amsmath}	
\usepackage{amssymb}	
\usepackage{float}
\usepackage{subcaption}
\usepackage{multicol}

\newcommand{\lya}{\,Lyman-$\alpha$ }

\newcommand{\angstrom}{\text{\normalfont\AA}}

\title[Lyman Alpha Spectral Recovery]{The Utility of Lyman-alpha Emission Lines as a Probe of Interactions between High Redshift Galaxies and their Environments}

\author[H. J. T. Childs et al.]{
Henry J. T. Childs,$^{1}$\thanks{Email: H.Childs@warwick.ac.uk}
Elizabeth R. Stanway,$^{1}$
\\
$^{1}$Department of Physics, University of Warwick, Gibbet Hill Road, Coventry, CV4 7AL, UK
}

\date{Accepted 2018 July 20. Received 2018 July 20; in original form 2017 August 07}

\pubyear{2018}

\begin{document}
\label{firstpage}
\pagerange{\pageref{firstpage}--\pageref{lastpage}}
\maketitle

\begin{abstract}
  The Lyman-$\alpha$ emission line is the strongest feature in the spectrum of most high redshift galaxies, and is typically observed as being highly asymmetric due to galactic outflows. Quantifying this asymmetry is challenging.  Here, we explore how measurements of one parameterisation, Lyman-$\alpha$ skewness, are affected by instrumental resolution and  detection signal-to-noise and thus whether this can be extended throughout the archive.  We model these effects through simulated lines and apply our derived corrections to existing archival data sets (including sources observed with FORS2 and DEIMOS) to reconstruct the intrinsic line emission parameters. We find a large uncertainty in parameter reconstruction at low resolutions ($R\le3000$) and high skew values, as well as substantial random errors resulting from the masking of sky lines. We suggest that interpretations of spectral line asymmetry should be made with caution, while a simpler parametrization, like B/R (blue-red flux asymmetry), is likely to yield more robust results. We see a possible weak trend in velocity width with mass, although there is no evidence in the data for a reliable correlation of skew with galaxy mass, star formation rate or age at $z=4-5$. Using our results, we investigate the possibilities of recovering emission line asymmetry with current and future instruments.
\end{abstract}



\section{Introduction}

The Lyman-$\mathrm{\alpha}$ emission line is produced by the electron recombination transition from the second to ground state of Hydrogen. Its emission occurs at $\mathrm{\lambda_{rest} = 1216\, \angstrom}$ and is common in star forming galaxies, triggered by strong  highly ionizing radiation from young hot stars.
Due to its  relative strength and its shift longwards of the atmospheric cut-off in observations above $\mathrm{z = 2}$, \lya has been extensively used as an diagnostic tool to find and investigate high redshift galaxies \citep[e.g.][]{hu96,2004ApJ...607..704S,matsuda04,ouchi08,nilsson09,yamada12}. \lya also yields insight into the escape fraction of highly energetic photons and hence the re-ionisation of the Universe \citep[e.g.][]{cowie09,verhamme15,verhamme16b,sobral16}. Galaxies observed with the \lya emission feature are aptly named Lyman-$\mathrm{\alpha}$ emitting galaxies (LAEs).

The spectral profile of \lya is shaped by the photon escape path in its local environment, due to its strong resonant scattering from neutral hydrogen in the interstellar medium (ISM) and circumgalactic medium (CGM). As a result, its emission reflects the nature of its host galaxy and the observed line of sight. The combination of these effects means that \lya can be observed as double peaked, asymmetric and sometimes exhibiting a P-Cygni profile \citep[e.g.][]{stern99,bunker00,2003MNRAS.342L..47B,shapley01,shapley03,verhamme06,steidel10,yamada12,verhamme16b}. The double peaked profile is produced by the emission going through an optically thick cloud of HI. The central wavelength of \lya is strongly absorbed and re-scattered. The scattered photons undergo a random walk, losing or gaining energy in the process until they escape resonance and are re-emitted in the blue and red wings of the line. If the blue wing is preferentially absorbed, or the red wing boosted by back-scattering from an expanding medium, this produces an asymmetric single line profile, a common feature seen in star-forming galaxies \citep[e.g.][]{verhamme06,ouchi08,steidel10,wisotzki16}. HI regions in the intergalactic medium (IGM) at different redshifts also create a series of \lya absorption features bluewards of the observed host \lya line, creating a \lya forest. If there is dense gas in the host or CGM, the limit of this process may  imprint additional asymmetry to the observed spectral line, as the bluewards line flux is absorbed  \citep{rauch97}. In order to infer details of the LAE galaxy environment, the line profile needs to be carefully examined and compared to radiative transfer models. This is typically achieved with measurements at high resolution and signal-to-noise (S/N) and thus is typically only possible at low redshift \citep[e.g.][]{verhamme06,dijkstra07b,dijkstra07}. However an overall measure of line aymmetry is more straightforward and confirms that most star forming galaxies in the distant Universe show a single asymmetric peak. 

A common picture of \lya radiative transfer in star-forming galaxies describes emitted photons as scattering off concentric spherical shells of gas, each ejected from the host galaxy due to supernova or stellar wind kinetic energy deposition and in outwards motion \citep{verhamme06}. This is consistent with observations which suggest that outflows are ubiquitous at high redshifts \citep[e.g.][]{adelberger03,pettini01,shapley03}. Precise measures of the quantity and state of the outflowing material require a range of rest UV spectral features, such as C IV ($\mathrm{\lambda_{rest} = 1549\, \angstrom}$) and thus high resolution and S/N data. It has been suggested that in $\mathrm{3<z<7}$ LAEs, a trend between asymmetry of the \lya line and axial ratio in rest UV-morphology may be linked to the detection of outflows \citep{u15}. A large spread in asymmetry, parametrized by skewness \citep[see][and equation \ref{eq:skgauss}]{mallery12}, was observed at low axial ratios. As the axial ratio increases the spread in skewness decreases. This has been interpreted as suggesting that the line asymmetry is caused by a galaxy-wide outflow, emitted along the polar axis of a disk-like galaxy. In this model, a galaxy seen face-on will have a low axial ratio and a high line asymmetry, while one seen edge-on will not. Such an interpretation presents an interesting possibility: if \lya emission and rest UV morphology can be used as independent and accurate diagnostics, large existing datasets could be used to provide statistics on outflows in the high redshift Universe. 

However, in addition to spectral effects caused by the ISM and intervening IGM, the effect of detection at a telescope also needs to be carefully investigated. By observing the effect of instrumental resolution and the inclusion of different peak intensity S/N values, the recoverability and reliability of \lya spectral properties can be investigated. This allows us to evaluate the potential information content of different datasets, at varying resolutions and S/N ratios. 

In this paper we look at the effect of a range of instrumental resolutions and peak intensity S/N on simulated spectral profiles of \lya. The effect of sky line masking and use on existing data sets is also discussed. The simulation details and data sample are described in section \ref{sec:sims}. In section \ref{sec:results} the results of the simulation are presented. In section \ref{sec:sky lines} the effects of the night sky emission spectrum and sky line masking on the practical measurement of sky line profiles are discussed. In section \ref{sec:discussion} the effects on asymmetry measurement of instrumental resolution and noise are discussed and real data applications are shown. We comment on implications of our analysis in the context of previously published work. A summary of our conclusions is presented in section \ref{sec:conclusions}.


\section{Simulations and Data Handling}\label{sec:sims}

To investigate the effect of instrumentation on the \lya emission profile we need a quantitative parametrisation of the spectral width and asymmetry of the emission line. For direct comparison with the inferred physical interpretation proposed by \citet{u15}, we characterize the \lya emission line profile with a skewed Gaussian model, adapted from \cite{mallery12}:

\begin{equation}
	\mathrm{Flux = A \, e^{-0.5((\lambda-x)/\omega)^{2}} \left( \int_{-\infty}^{\mathrm{s}(\lambda-x)/\omega}e^{-t^2/2} \, dt \right) + c}
	\label{eq:skgauss}
\end{equation}

By fitting this equation to data or simulated data we can recover the values of flux normalization (A), central wavelength ($\lambda$), standard deviation ($\mathrm{\omega}$) and skew (s). In this scheme the value of skew dictates the asymmetry in the emission line. A positive skew indicates a boosted red wing and suppressed blue wing, while a negative skew reverses this. For a zero skew the equation reverts to a simple symmetric Gaussian form.
This simple model of \lya emission lines assumes a constant continuum level (c) as well as ignoring the absorption component of P-Cygni profiles. This simplification is justified by typical line profiles seen at redshifts $z>3$ \citep{mallery12}.

\begin{figure*}
  \centering
  \begin{minipage}[b]{0.98\textwidth}
    \includegraphics[width=\textwidth]{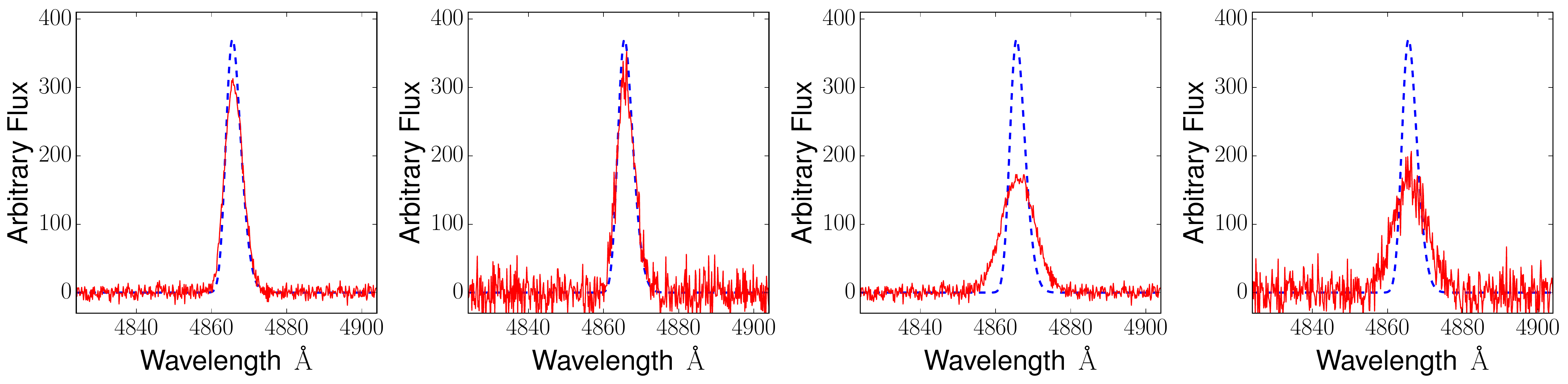}
    \caption{Example of simulated skewed \lya emission lines and their subsequent detection at combinations of resolution and S/N of (3000,30),(3000,10),(1000,30) and (1000,10) respectively (from left to right). In blue is the intrinsic emission line profile with skew of 2 and FWHM of 7\,\AA. In red is the measured profile.}
    \label{fig:measprof}
  \end{minipage}
  \hfill
  \begin{minipage}[b]{0.98\textwidth}
    \includegraphics[width=\textwidth]{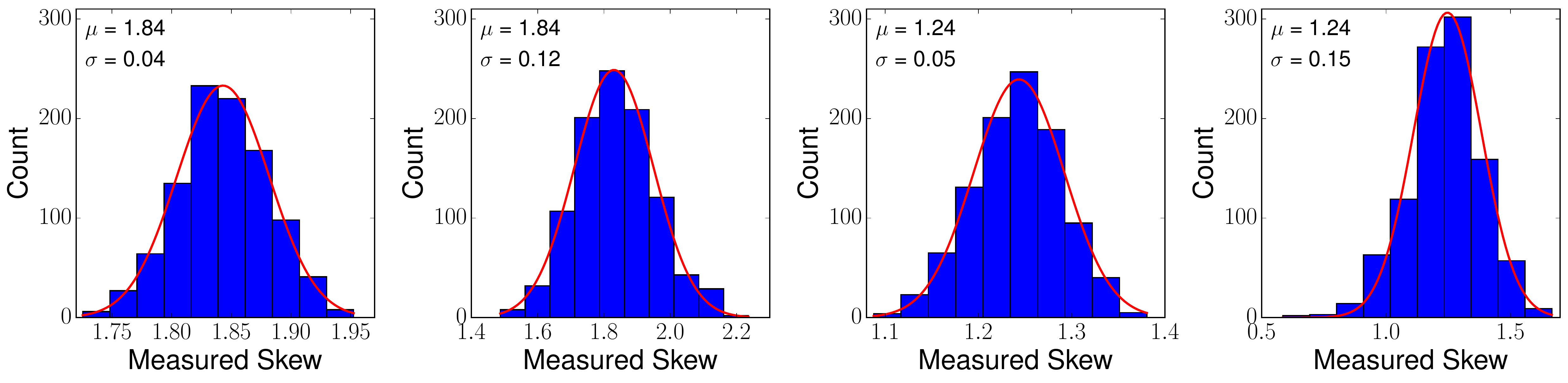}
    \caption{Output of 1000 noise iterations at intrinsic (i.e. simulation input) skew of 2 and FWHM of 7\,\AA\ for the resolution and S/N combinations seen in figure \ref{fig:measprof}. A Gaussian fit to the data is shown with a solid red line, with measured distribution mean and standard deviation annotated in each panel.}
    \label{fig:nsit}
  \end{minipage}
\end{figure*}

We use  equation \ref{eq:skgauss} to model the recovery of simulated emission lines with known intrinsic FWHM and skew parameters, after accounting for detector resolution and peak line S/N.  For the purposes of the simulation, a fixed redshift $\mathrm{z=3}$ is chosen. This redshift implies a \lya detection at $\mathrm{\lambda =4864\, \angstrom}$, in the SDSS g band. 
We restrict intrinsic FWHM to $\mathrm{1-7 \, \angstrom}$, giving a spectral velocity width range of $\mathrm{75-500 \, km\,s^{-1}}$, motivated by velocities that are physically seen in star-forming regions \citep[e.g.][]{2014MNRAS.439.1101Z}. The intrinsic skew was restricted to values of $\mathrm{0-6}$, motivated by the range seen in the large sample presented by \cite{u15} and \citet{mallery12}.

With this sample of emission lines, the effects of detection can be simulated. First we convolve the profiles with a range of instrumental resolutions, motivated by current and future instrumentation: $R$\,=\,300,\,1000,\,3000,\,10000. We sample our simulated lines using a pixel scale of $\mathrm{0.5\,\angstrom}$. The pixel width was chosen to ensure that at all times the instrumental resolution is fully sampled, so as not to introduce further ambiguity in line fitting. Although the effect of pixel scale on resulting errors should be taken with caution, as we discuss in section \ref{sec:binning}.
We add Gaussian noise to the convolved and re-sampled skewed Gaussians to give a range of S/N measurements typical of  real data: S/N\,=\,3,\,10,\,30,\,100. This signal to noise is defined in the pixel at the peak of the emission line, and the true S/N will be lower in the wings of the lines or on the continuum.

It is important to note that a random Gaussian noise model was used as the primary noise characteristic in this work. This is the appropriate noise level if dominated by residuals from sky background subtractions. However it will underestimate the noise in strong lines, where the Poisson uncertainty due to the number of detected electrons in each pixel can also be an important component of the overall noise budget. This will preferentially affect the wings of the lines, in which the fractional uncertainty is highest. As a test, Poisson noise was also calculated for each spectra, and was found to be consistently below the Gaussian noise level. It typically contributed less than 10 per cent of the noise budget. At very low values of total S/N, i.e. below 10, the Poisson noise contribution unsurprisingly becomes dominant, reaching 30 per cent in extreme cases. We note that we do not present results in these regimes, since at very low S/N, even including only the Gaussian noise sky background component, the line parameters are already obscured by the noise, as section \ref{sec:results} demonstrates.

The generated profiles are then measured by fitting equation~\ref{eq:skgauss} to the results using a least squares algorithm. To evaluate statistical uncertainties, we run each combination of skewed Gaussian, resolution and S/N over 1000 noise iterations. The combined noise iteration results are then themselves fit by a Gaussian distribution where the measured skew and FWHM, their means and standard deviations, are calculated. An example of this process for producing simulated emission lines and artificially detecting them with realistic instrumental limitations is illustrated in figures \ref{fig:measprof}-\ref{fig:nsit} for a specific FWHM and skew combination, across a small range of resolution and S/N.

As a final step, we convert all FWHM results into velocity widths. As both velocity width and resolution scale $\mathrm{\propto(1+z)}$ all results presented are valid for any choice of redshift, assuming the measured skewed Gaussian is well sampled. By following this procedure, simulating the entire range of parameters, we observe how the intrinsic spectral properties are transformed by instrumental effects, and whether they can be reliably recovered. 

We define the properties of a line as {\it recoverable} if it has a calculated intrinsic parameter uncertainty $\mathrm{\sigma_{int} < 100\%}$ (i.e. the derived uncertainty is less than the derived value) and is not degenerate. We note that this allows generously large errors and many science applications would require smaller uncertainties. We define a property as {\it unrecoverable} if there is no clear solution for the relation between intrinsic and observed properties in our simulations, or if the associated uncertainty exceeds the criterion above.


\section{Results}\label{sec:results}

\begin{figure*}
\centering
\includegraphics[scale=0.75]{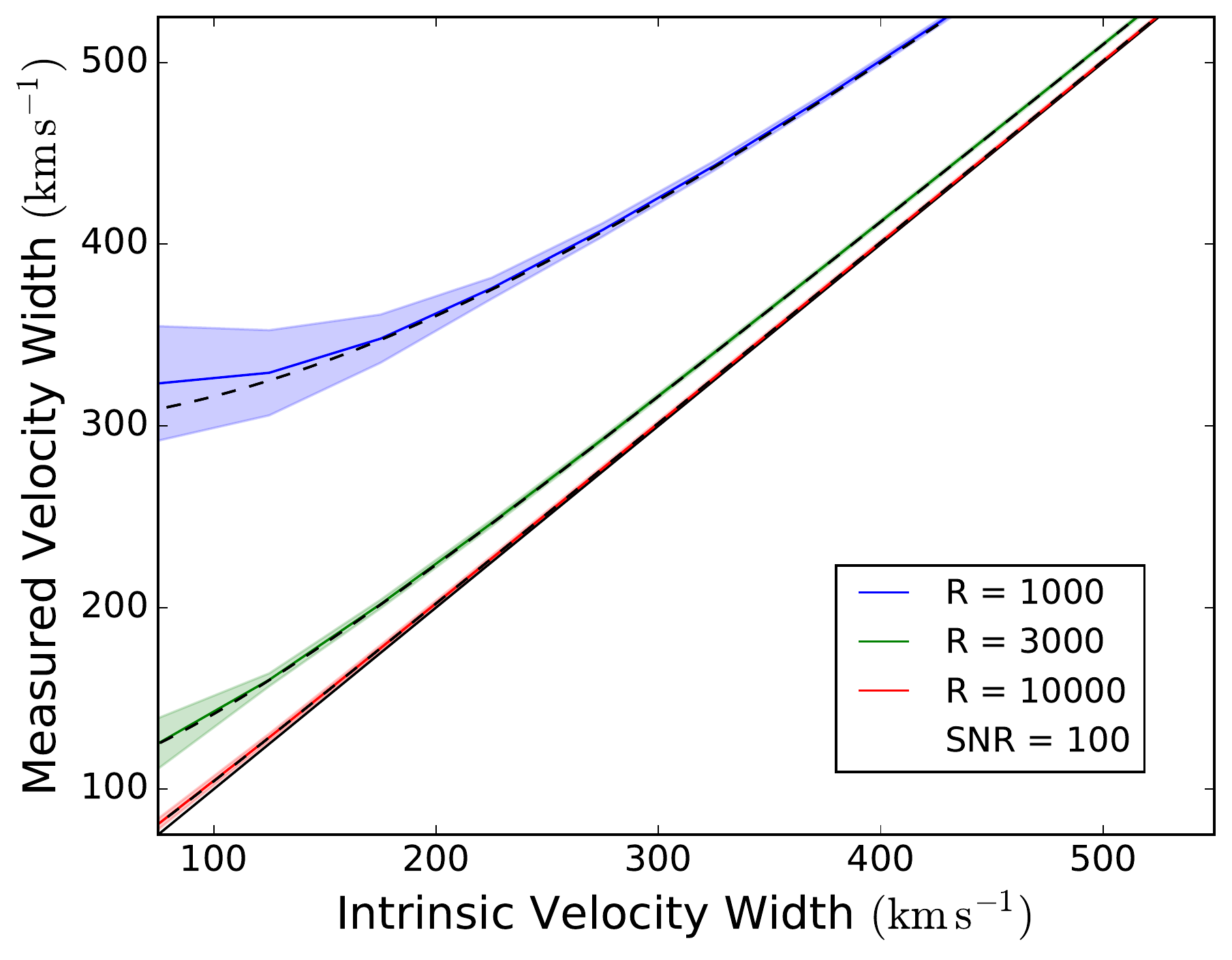}
\caption{Measured vs. intrinsic velocity width seen with $\mathrm{S/N=100}$, at intrinsic $\mathrm{skew = 6.0}$ and at three sample resolutions. The black dashed line shows the expected measured velocity width of a pure Gaussian function at each resolution. Shaded regions indicate the random uncertainty on the measured value based on 1000 noise iterations. If the intrinsic velocity width were unaffected by instrumental resolution and noise the measured velocity width would perfectly track the solid black line.}
\label{fig:sktrk100}

\centering
\includegraphics[scale=0.75]{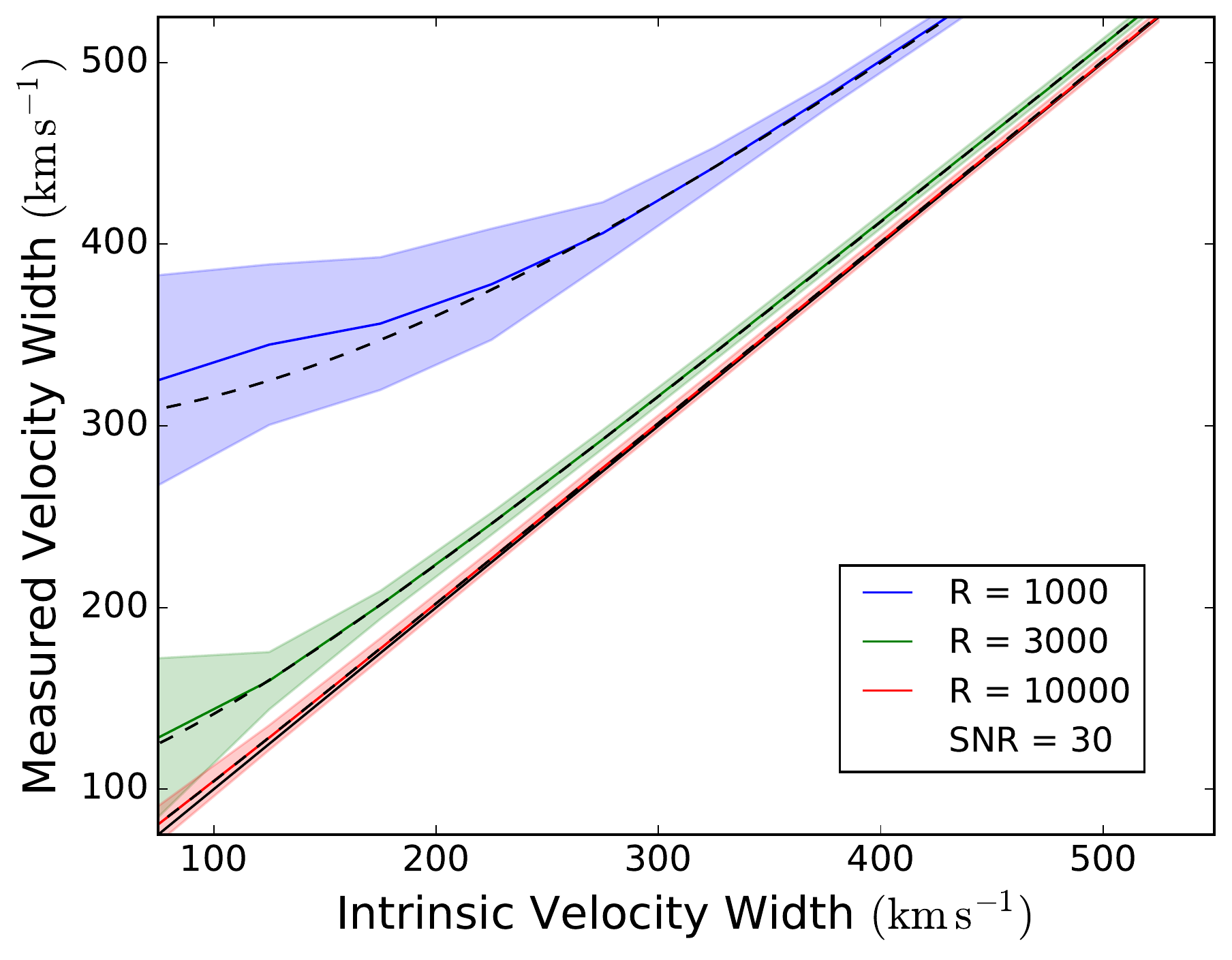}
\caption{As in figure \ref{fig:sktrk100} but with $\mathrm{S/N=30}$.}
\label{fig:sktrk30}
\end{figure*}

\begin{figure*}
\centering
\includegraphics[scale=0.75]{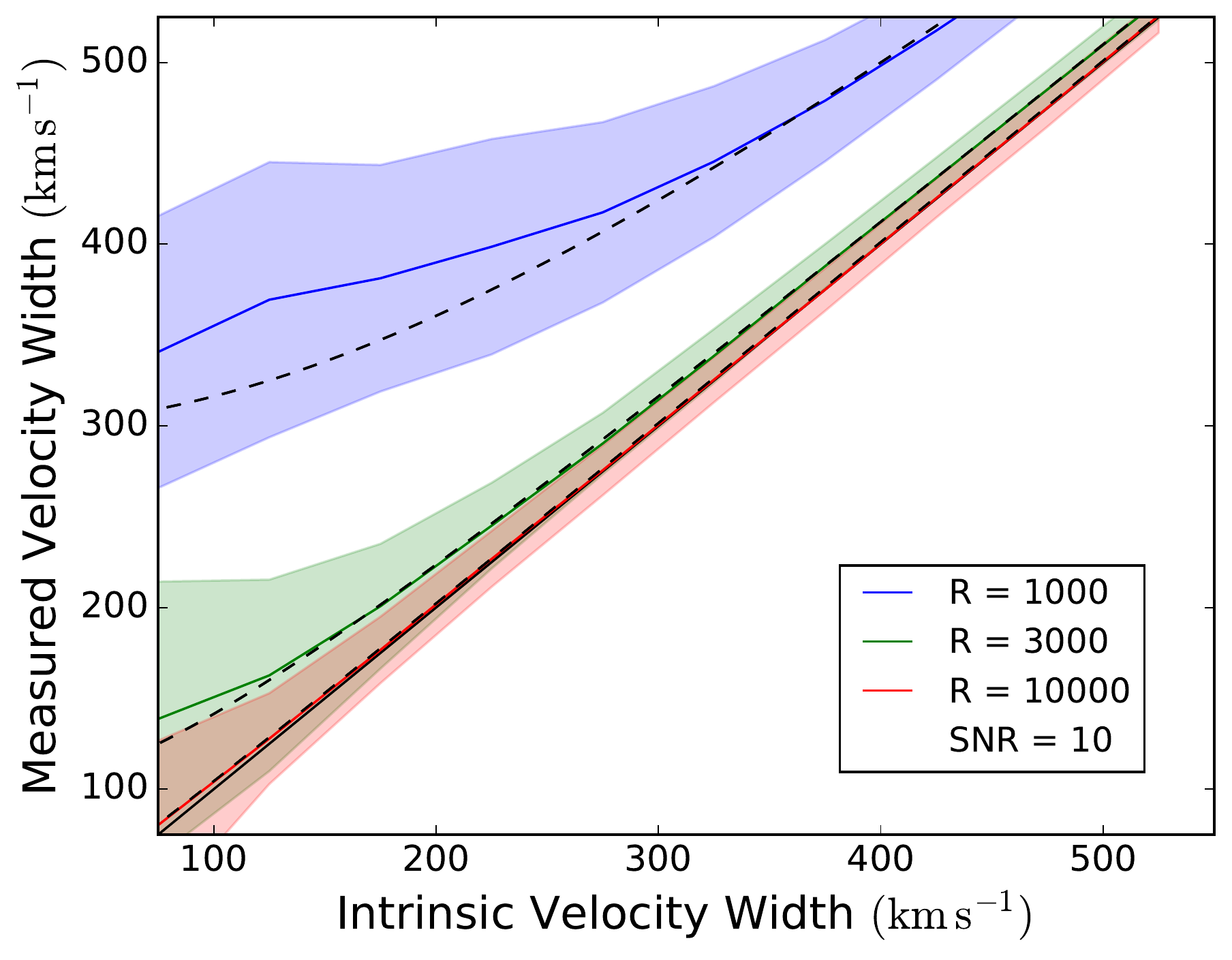}
\caption{As in figure \ref{fig:sktrk100} but with $\mathrm{S/N=10}$.}
\label{fig:sktrk10}

\centering
\includegraphics[scale=0.75]{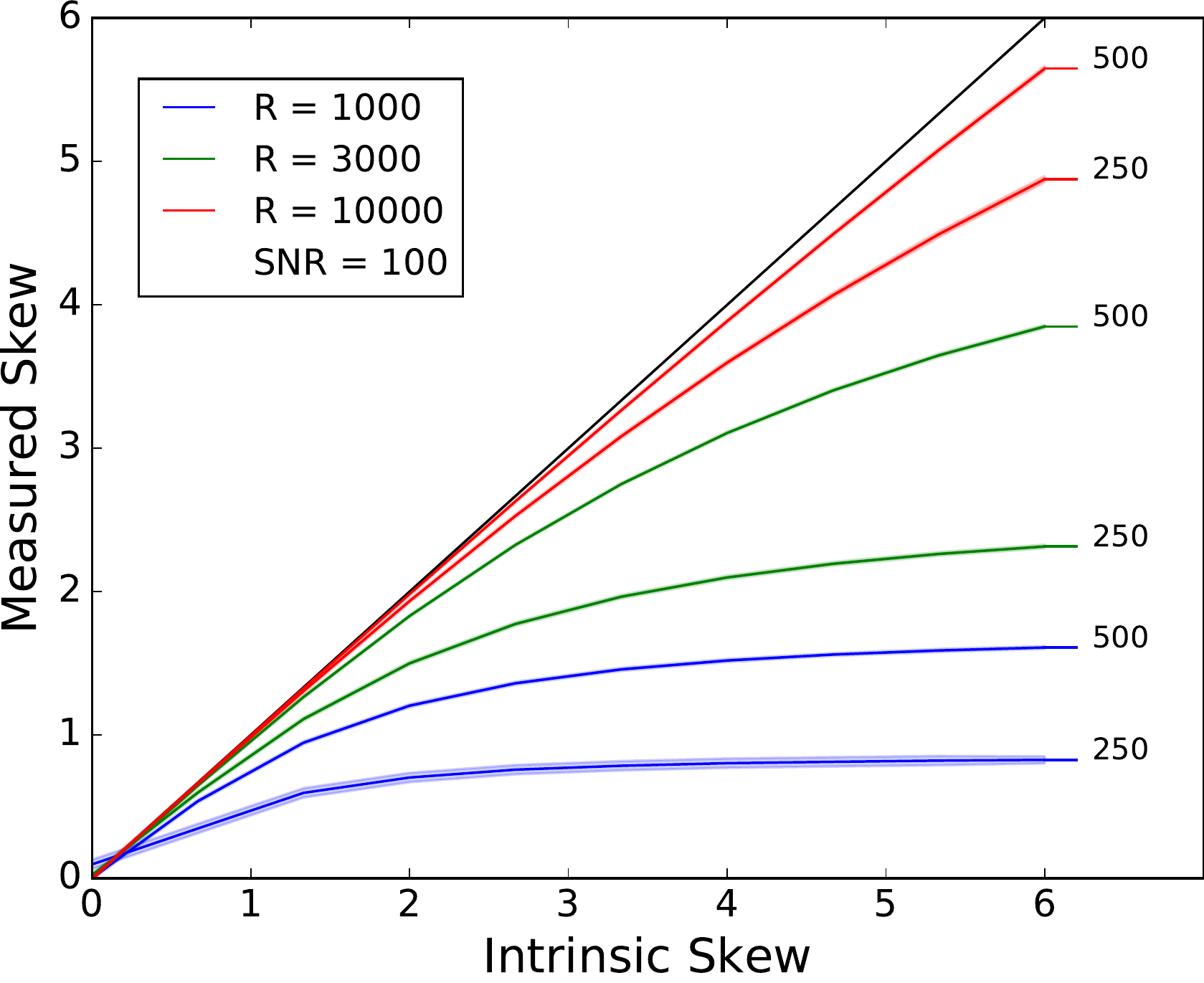}
\caption{Measured vs. intrinsic skew seen at different intrinsic velocity widths, labelled on the right-hand side in $\mathrm{km \,s^{-1}}$. Measurements are at $\mathrm{S/N=100}$, and at three sample resolutions. As before, shaded regions indicate random uncertainty from 1000 noise iterations. If the intrinsic skew were unaffected by instrumental resolution and noise the measured skew would perfectly track the solid black line.}
\label{fig:veltrk100}
\end{figure*}

\begin{figure*}
\centering
\includegraphics[scale=0.75]{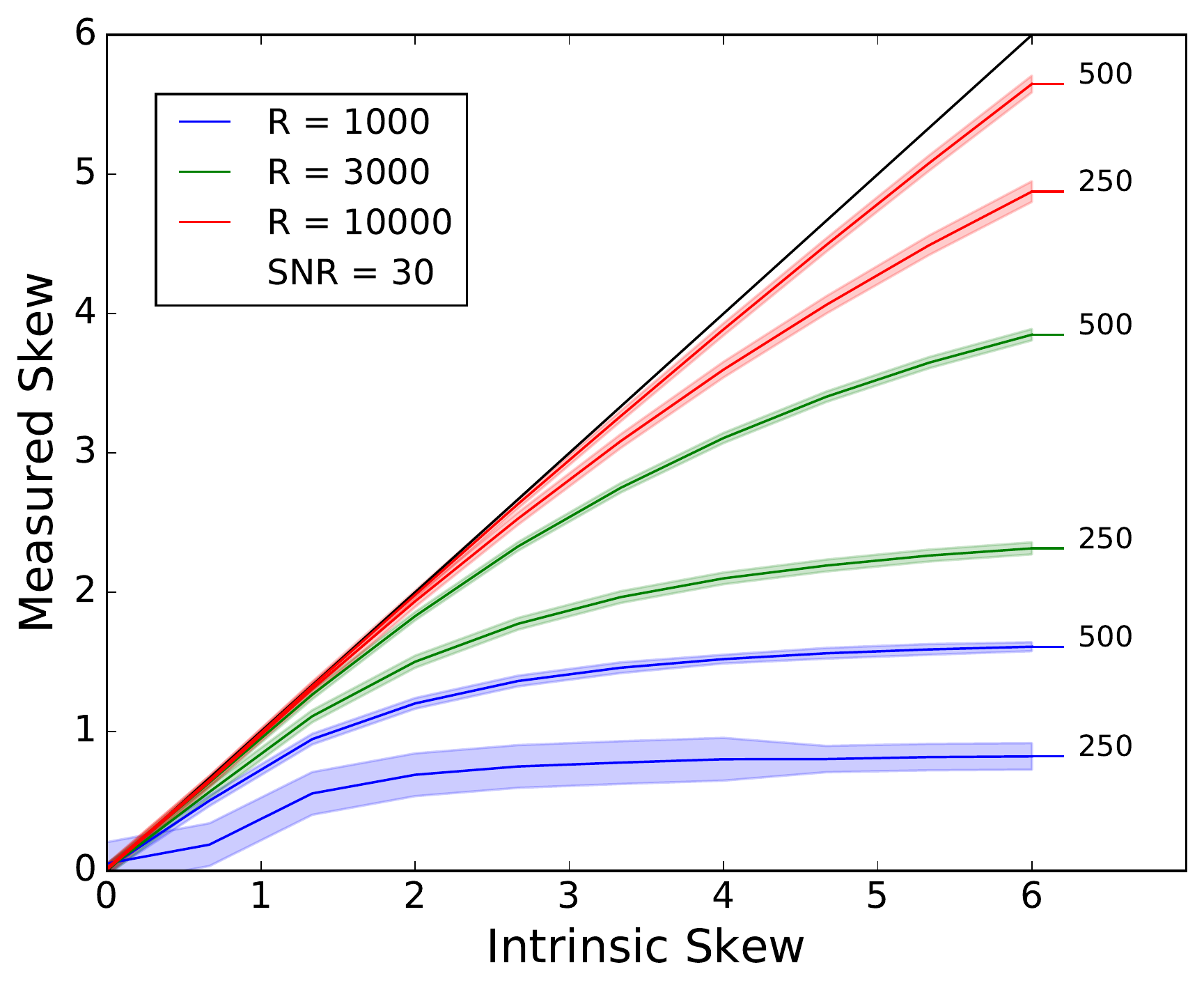}
\caption{As in figure \ref{fig:veltrk100} but with $\mathrm{S/N=30}$.}
\label{fig:veltrk30}

\centering
\includegraphics[scale=0.75]{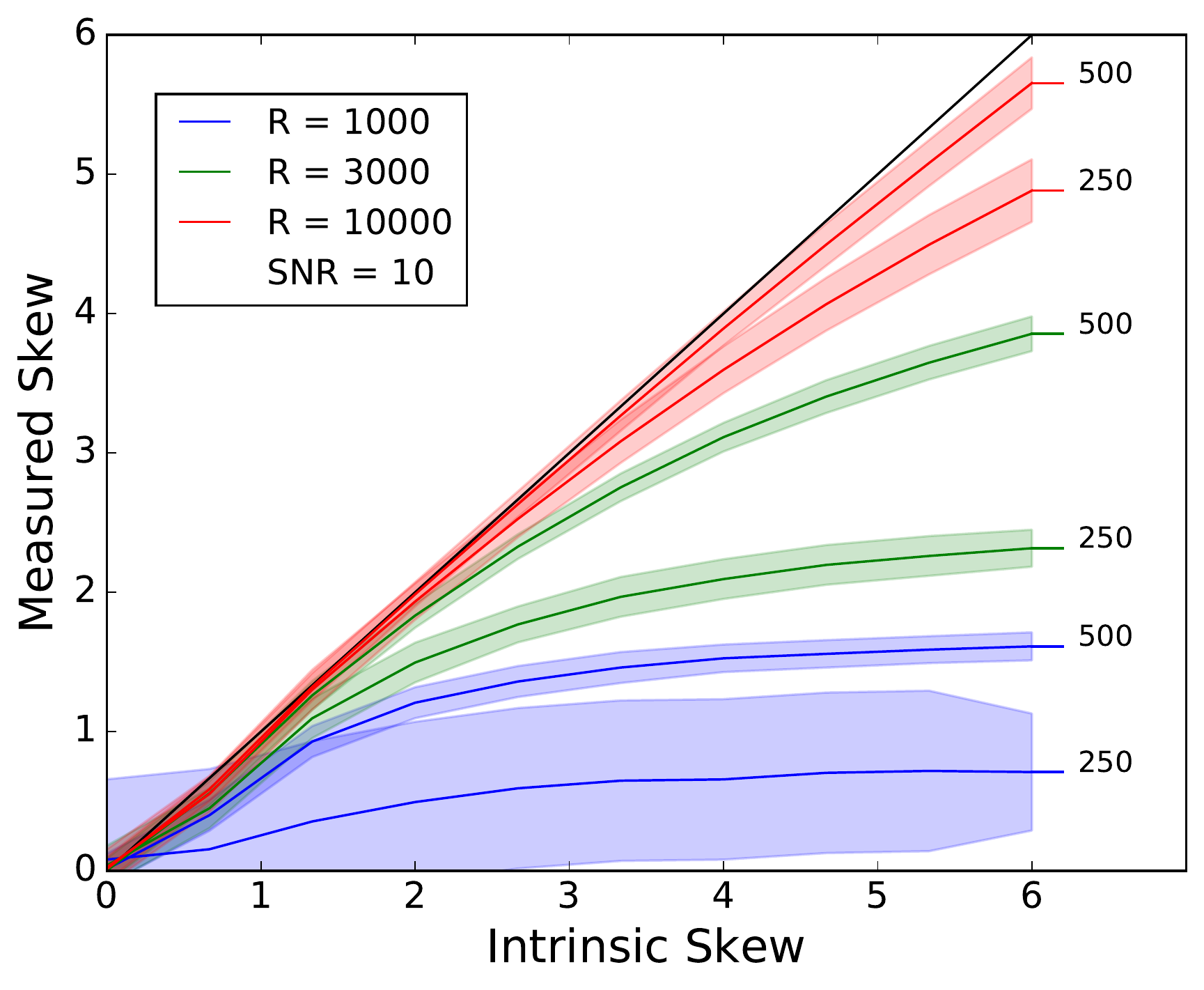}
\caption{As in figure \ref{fig:veltrk100} but with $\mathrm{S/N=10}$.}
\label{fig:veltrk10}
\end{figure*}

\begin{figure*}
\centering
\includegraphics[scale=0.75]{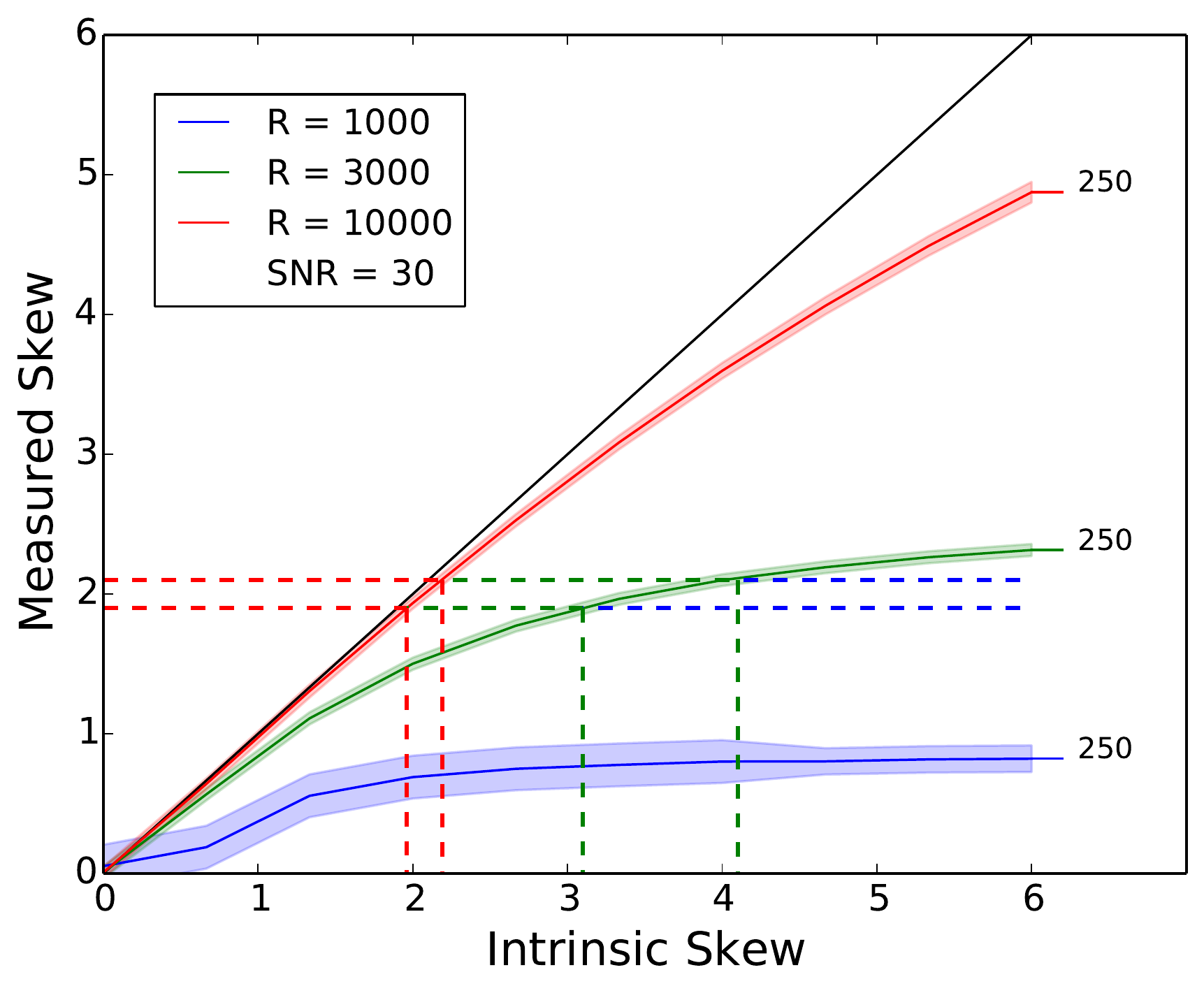}
\caption{Example recovery of an intrinsic line skew from the measured value. Given a measured line skew of $\mathrm{sk_{meas} = 2.0 \pm 0.1}$ at $\mathrm{S/N=30}$ and with $\mathrm{\Delta v_{int} = 250 km\,s^{-1}}$, we demonstrate the implied intrinsic skew in the cases where $\mathrm{R = 1000,\,3000,\,10000}$. In each case the recovered intrinsic skew corresponding to the resolution is indicated by dashed lines in a matching colour. Intrinsic skew values of sk$_\mathrm{int}$\,=\,unknown,\,$3.6 \pm 0.5$,\,$2.1 \pm 0.1$ were recovered at the three resolutions, suggesting that an accurate measurement is only possible at the highest resolution. Where no intersection with the recovery line is found the skew is deemed unrecoverable, as is seen at the lowest resolution.}
\label{fig:veltrkrec}
\end{figure*}

Results of the simulations are presented in figures \ref{fig:sktrk100}-\ref{fig:veltrk10}. The results at R = 300 and S/N = 3 showed no recoverability in skew. The entire range of input parameters produce degenerate results, with measured values dominated by the instrumental resolution and noise. As a result no recoverability plots are presented for $\mathrm{R = 300}$ or $\mathrm{S/N = 3}$.

Figures \ref{fig:sktrk100}-\ref{fig:sktrk10} show that the velocity width is recoverable over the remaining range of resolution and S/N. We also observe that velocity width recovery is skew-independent. The offsets measured from the simulated data follow what is expected of an intrinsic Gaussian line profile being convolved with another Gaussian kernel modelling the instrumental resolution. When the intrinsic velocity width is below the width of the resolution kernel, the measured value is dominated by the resolution width and the uncertainty on the reconstructed line width is very large.

At resolution $\mathrm{R = 10000}$ we see offsets between the input and measured velocity width at  $\mathrm{< 5.1\,km\,s^{-1}}$ or $\mathrm{\sim 5\%}$. At resolution $\mathrm{R = 3000}$ much larger offsets are seen, up to $\mathrm{40\, km\,s^{-1}}$, but these remain consistent with the expectation when deconvolved with the instrumental resolution. $\mathrm{R = 1000}$, as expected, exhibits the largest offsets in measured results, but now also shows an apparent deviation from the expected Gaussian convolution curve which is most pronounced at higher noise levels. Across all resolutions and S/N the velocity width is seen to be recoverable through a combination of deconvolution from the instrumental width and correction for any systematic offset, with no degeneracies observed in measured-intrinsic values.

The recovery of skew, shown in figures \ref{fig:veltrk100}-\ref{fig:veltrk10}, is observed to have strong dependence on resolution, velocity width and S/N. At $\mathrm{R=10000}$ the values of intrinsic skew are easily recovered across all sampled intrinsic velocity widths and S/N values. While the measured lines show a systematic offset from the input intrinsic values, this can be modelled and a correction applied. With current instrumentation, resolutions this large are often only available in echelle spectroscopy. At $z>3$, a typical $L^\ast$ object will have magnitudes fainter than 24.5 \citep{shapley03}. Given normal observing conditions such an object would require exposure times $>>$ 10\, hours to reach a signal to noise on the continuum of three per resolution element on an 8m telescope\footnote{Estimated using the ESO FORS exposure time calculator, assuming a 1 arcsecond wide slit, typical observing conditions, dark time and the 600RI (R$\sim$1000) grating of the FORS2 instrument.}. While the peak of the line will have a substantially larger signal-to-noise this will still be a challenging observation, most $z>3$ Lyman $\alpha$ and Lyman break surveys have been carried out at much lower resolution for this reason \citep[e.g. $R\sim300$][]{shapley03}.

At lower resolutions we see that measured skew is suppressed relative to the input, and measurement error increases. At $\mathrm{R=1000}$ and relatively narrow lines with FWHM$=250$\,km\,s$^{-1}$, we observe a limit in intrinsic skew of $\mathrm{sk\sim1.0}$, above which the measured skew becomes constant and the reconstruction of the intrinsic skew is degenerate. For broader lines (FWHM$=500$\,km\,s$^{-1}$) the measured skew still becomes degenerate with intrinsic skew, but this happens at slightly higher line asymmetries. Likewise at an intermediate resolution, R=3000 and for narrow lines we see measured skew saturating when intrinsic skew sk$\ga3$. The overall effect seen by the simulation of instrumental resolution is a systematic underestimation of intrinsic skew, with random uncertainties on this systematic offset set by signal-to-noise ratio.

As well as degeneracy in measured-intrinsic values, the uncertainty set by the width of the measured skew distribution (e.g. figure \ref{fig:nsit}) also contributes to the inability to recover intrinsic values successfully. While in the simulated case, it is possible to run 1000 noise iterations to explore this distribution, in real data we will have only one realisation and must estimate the effect of this uncertainty. By taking an example measurement we can explore the precision to which a value can be recovered, given this additional uncertainty. In this example, see figure \ref{fig:veltrkrec}, we choose a typical measured value of skew $\mathrm{sk_{meas} = 2.0 \pm 0.1}$ and intrinsic velocity width $\mathrm{\Delta v = 250\,km\,s^{-1}}$ with a $\mathrm{S/N = 30}$. At $\mathrm{R = 1000,3000,10000}$ we find recovered values of $\mathrm{sk_{int} = unknown, 3.6 \pm 0.5, 2.1 \pm 0.1}$ respectively, where the lowest resolution value is not physically meaningful in our model. For values that can be recovered, an increase in uncertainty is seen in all results - i.e. the recovered intrinsic value is less well constrained than the measured value.


\section{Sky line Masking}\label{sec:sky lines}

In addition to instrumental resolution and noise, another important feature to consider in real world ground-based \lya detection is the narrow night sky OH emission line background. At the red end of the optical spectrum blanketing from sky lines becomes a large potential contaminant \citep{osterbrock96}. While sky lines are processed and removed from observational data, the photon noise residuals inevitably associated with large measured fluxes are likely to alter any measured characteristic emission line parameters. To consider the possible effect of sky line masking we first generate an emission line  with FWHM and skew of $\mathrm{500\, km\,s^{-1}}$ and 1.0 respectively, and convolve with an instrumental resolution of $\mathrm{R = 3000}$. Then by assuming the night sky emission is effectively at rest with respect to the spectrum, we can set the night sky emission width $\mathrm{\sigma}$ as purely determined by the instrumental resolution. We simulate the effects of a sky line nearby by masking and excluding a wavelength range $\pm\mathrm{\sigma}$ (sky) from the fit of the generated spectrum. We then measure the FWHM and skew from the masked spectrum. 

\begin{figure}
\includegraphics[scale=0.4]{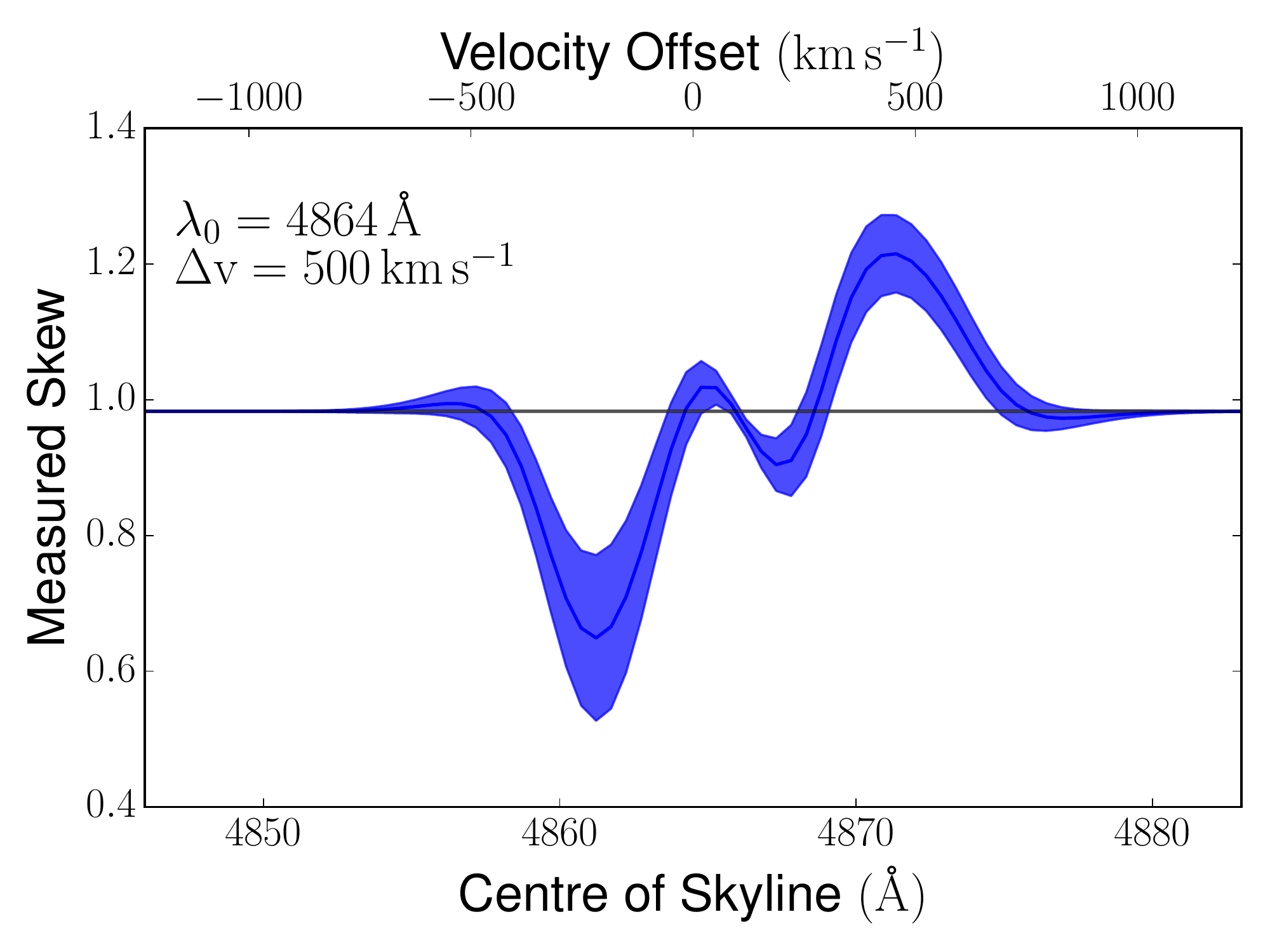}
\caption{Effects of a sky line mask exclusion ($\mathrm{R=3000}$) being traced along a skewed Gaussian line, with a central wavelength $\mathrm{\lambda_0 = 4864\,\angstrom}$.}
\label{fig:slnrmv}
\end{figure}

Figure \ref{fig:slnrmv} shows the result of masking across sequential regions of the example emission line. As the sky line mask is shifted across the emission line we first see the usual offset in measured skew due to instrumental resolution as seen in section \ref{sec:results}, where the measured skew is suppressed relative to the intrinsic skew. As the sky line mask enters the emission line strong deviations from the expected measurement at the given resolution are observed. An overall reduction in skew is seen when the sky line peak is bluewards of the line center and an increase is seen redwards of the line centre. At its peak, a deviation of $\sim40\%$ is observed. This suggests caution should be applied when interpreting skewed line profiles which lie close to sky lines subtractions. In these cases, the measured value is extremely sensitive to proximity to the line, and the intrinsic profile. While the sky line centroid is well known and can be refined in the data itself, the intrinsic line properties are impossible to know {\it a priori}. The increased uncertainty in the intrinsic skew can exceed 100 percent of the measured value, which for most applications makes the line fit essentially useless. This means caution and careful error analysis needs to be enacted when line features are spectrally coincident with sky lines. We note that this example is for an intrinsically broad line and relatively high resolution - a scenario in which the uncertainties on line recoverability should be minimised. The effects are likely to be more pronounced for narrow lines and lower spectral resolutions. In fact, this will need to be modelled for each instrument setting used, and may render some regions of redshift, and velocity-width parameter space essentially unaccessible.

We extend this analysis  by taking the expected typical sky spectra, obtained from relevant exposure time calculators, from several specific instruments and sites. We observe the effect that the addition of this sky line background has on the measured skewness of a \lya line of varying luminosity. We take an example emission line of fixed velocity width $\mathrm{\Delta v = 250 km\,s^{-1}}$, skew $\mathrm{sk = 2.0}$ and  luminosities log(L/erg\,s$^{-1}$) = 41,\,42,\,43. We take as examples the FORS2, DEIMOS and forthcoming EAGLE instrument, which have resolutions 660, 3000 and 10000 respectively. To account for the usage patterns and relative sensitivities, we calculate models assuming exposure times of 5 hrs, 3.5 hrs, 10 mins respectively.

\begin{figure*}
\centering
\includegraphics[scale=0.75]{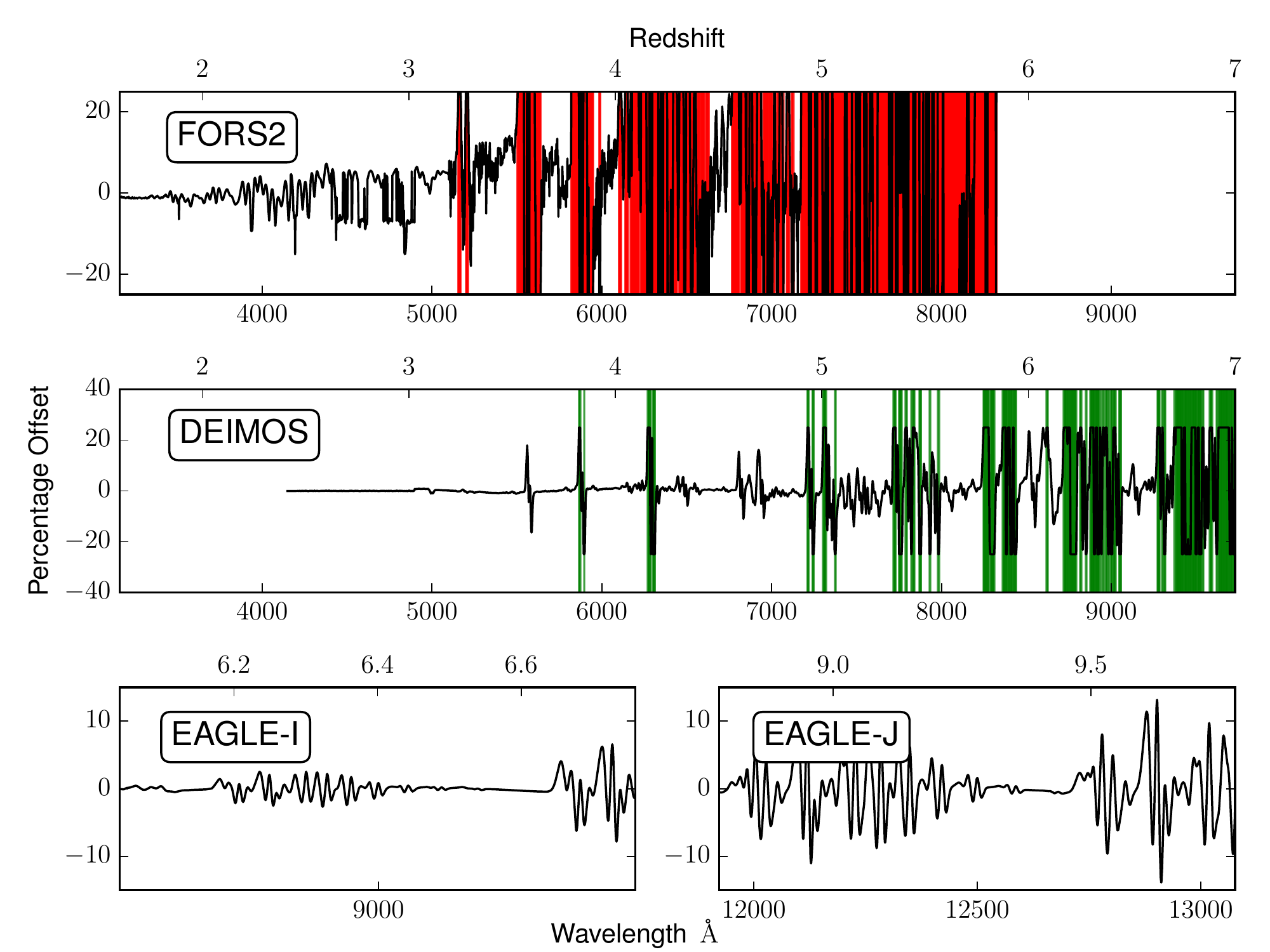}
\caption{Example percentage offsets in measured skew, relative to that measured in the absence of sky line emission, for the FORS2, DEIMOS and EAGLE instruments, for an object of line luminosity log(L$_\mathrm{Ly\alpha}$/erg s$^{-1}$)=42. Exposure times of 5 hrs, 3.5 hrs and 10 mins were used respectively to reproduce reasonable integration time cases for each facility. Shaded  regions indicate where the offset exceeds 25 per cent.}
\label{fig:full_skyline}
\end{figure*}

We explore the wavelength- and resolution-dependent effects of the sky line spectrum by stepping the central wavelength of our modelled emission line at 1\AA\ intervals from the lower to upper wavelength cutoffs appropriate to each instrument. This corresponds to exploring a series of lines across a range of redshifts in each case. We then derive the uncertainty associated with the measured skews as shown above, and plot this as a function of wavelength (or redshift) in figure \ref{fig:full_skyline}. As is apparent, a low resolution instrument such as FORS2 has very little usable spectral parameter space above $z=4$, with narrow windows opening around $z\sim4.5$ and $z\sim4.85$. At higher redshifts, the uncertainty on the measured skew exceeds 25 per cent. With a higher resolution instrument such as DEIMOS, the usable redshift range is much larger, with only regions dense in sky lines rendered unusable, although beyond $z\sim6$ recovering Lyman-$\alpha$ skew will be challenging even with this instrument. Windows of high recoverability exist between sky line complexes, notably at $5.07<z<5.35$,  $5.57<z<5.78$, $5.95<z<6.09$ and  $6.45<z<6.62$. By contrast, the much higher resolution EAGLE instrument under construction for the European Extremely Large Telescope (E-ELT) will not suffer substantially from sky line blanketing but will have a limited redshift range, and is also likely to have a much lower signal-to-noise per pixel, than the existing instruments. We note that none of the instruments modelled here will be useful in the range $z\sim6.8-8.7$.

Our criterion for recoverability in section \ref{sec:results} permitted a 100\% uncertainty on the recovered value. For physical interpretation, this is far too loose. In table \ref{table:skytable} we consider a tighter constraint, calculating the percentage of the spectral range covered by each instrument in the specified setting (i.e. the fraction of the redshift range) which would allow recovery of the lines with a fractional uncertainty on the line skew below thresholds of 0.1, 0.25, 0.4, 0.75 and 1.0. As expected, we see that the useable fraction of the spectrum strongly scales with the luminosity of the emission line (and hence its signal-to-noise in a fixed time integration). We note that even in the very short, ten minute integrations modelled, the EAGLE instrument will be excellent for the recoverability of line skews, although these may have to be for systems with a systemic redshift already well known.

The large uncertainties calculated above suggest that recovering intrinsic \lya skew parameters from ground based and low resolution instruments with uncertainties $<25$\% is possible for only about half the spectral range, see table \ref{table:skytable}, and becomes virtually impossible in the red where sky lines become frequent and strong. The exception occurs when working at sufficiently high spectral resolution to observe between sky lines, as DEIMOS does over much of the optical, and EAGLE certainly will in the near-infrared. Lower resolution instruments such as FORS2 do allow well constrained measures of line skew  when the source redshift  places \lya emission conveniently far from a sky line centroid. 

\begin{table*}
\centering
\caption{Percentage of each given instrument spectral range that is free of sky line contamination which exceeds a range of limiting fractional uncertainty thresholds, for exposure times matching those in figure \ref{fig:full_skyline}, at each given luminosity. Each column represents a different threshold in terms of fractional uncertainty on the measured skew value, while each row represents a different instrument and line luminosity combination. The example \lya emission line had an input skew sk=2.0 and velocity width $\mathrm{\Delta v = 250 km\,s^{-1}}$.}
\label{table:skytable} 
\begin{tabular}{|c|c|c|c|c|c|c|}
 \hline
   & & \multicolumn{5}{c}{Fractional Uncertainty Threshold} \\
  & & 0.1 & 0.25 & 0.50 & 0.75 & 1.0 \\ \hline \hline
 Instrument & log(L$_{\mathrm{Ly}\alpha}$/$\mathrm{erg\,s^{-1}}$) & \multicolumn{5}{c}{Percentage Usable} \\ \hline
 {\bf FORS2} & 41 &37 &47&54&63&74\\ 
 $1.5<z<5.7$ & 42 &38&52&69&71&77 \\
 R$\sim660$, 5 hrs & 43 &38&75&82&92&95 \\ \hline
 
 {\bf DEIMOS} & 41 &52&55&61&68&76\\
 $2.20<z<7.22$ & 42 &75&83&90&93&95\\
 R$\sim3000$, 3.5 hrs & 43 &98&100&100&100&100\\ \hline
 
  {\bf EAGLE (I band)} & 41 & 99 & 100 & 100 & 100 & 100 \\
$6.02<z<6.77$ & 42 & 100  & 100 & 100  &  100 &  100 \\
 R$\sim10000$, 10 mins & 43 &  100 &  100 &  100 & 100  & 100 \\ \hline
 
  {\bf EAGLE (J band)} & 41 & 70 & 81 & 89 & 93 & 96\\
$8.75<z<9.80$ & 42 & 98 & 100 & 100 & 100 & 100\\
 R$\sim10000$, 10 mins & 43 & 99 & 100 & 100 & 100 & 100 \\ \hline 
\end{tabular}
\end{table*}


\section{Discussion}\label{sec:discussion}

To be able to use \lya skew as an outflow diagnostic, \cite{u15} suggest that a trend exists between the scatter in skew and rest UV morphology. While they used high quality, $\mathrm{R\sim3000}$, data, we want to investigate whether the same diagnostic may be reliable for lower resolution data. As shown in section \ref{sec:results}, the accurate measurement and recovery of intrinsic skew is non-trivial. At large values of both instrumental resolution and galaxy velocity width, the values of skew can be reliably recovered through correction of the systematic offset between measured and intrinsic skew. However, poorer instrumental characteristics induce large scatters in the measured skew, particularly at low S/N. Most notably at $\mathrm{R=300}$ and $\mathrm{R=1000}$, we see almost no recoverability in the skew parameter space since the intrinsic skews are degenerate for a given measured skew. 

Overall, the effect of finite instrumental resolution is to reduce the measured value of skew relative to the intrinsic value. This may suppress the scatter in a distribution of measurements when compared to the scatter in the source distribution. When applied to real data this could cause significant difficulties in interpretation, and result in measurements with low apparent asymmetries. 

\subsection{Binning Data}\label{sec:binning}

Spectra from real data are often binned and smoothed, in-order to increase S/N. While this can allow fainter signals to be analyzed, the effect this may have on measured skew needs to be quantized. Previous work has suggested that excessive binning (i.e. a bin width greater than 0.5\,\AA) is likely to affect measurements of intrinsic line parameters \citep{robertson17}. By taking an example \lya emission line (skew = 2.0, $\mathrm{\Delta v = 250 km\,s^{-1}}$ and at z = 3) we can alter the sample binning and observe how this affects the overall ability to characterize its skew. We take this typical binned emission line and perform 1000 noise iterations over a small range of resolution and S/N to obtain the error on measurement. We compare this to previous, 0.5\,\AA\ pixel resolution, data to show the increase in percentage error caused by binning the data, as shown in table \ref{table:bintable}. 
 
We see a moderate increase in uncertainty, with the additional error term (added in quadrature to other factors) reaching a maximum of 47 per cent, but the increasing bin width has very little effect at low resolutions, and only becomes severe at high resolution at the largest bin sizes. As binning is usually done in conjunction with low S/N data we see that excessive binning, as expected, yields unrecoverable skews.

\begin{table}
\centering
\caption{Percentage uncertainty on measured skew as a function of the sample bin width of a skewed emission line with known intrinsic skew. Example \lya emission line has an input skew sk=2.0 and velocity width $\mathrm{\Delta v = 250 km\,s^{-1}}$.}
\label{table:bintable}
\begin{tabular}{|c|c|c|c|}
 \hline
 Resolution & S/N & Bin Width ({{\AA}}) & Percentage Uncertainty \\ \hline \hline
   & & 0.5 & 6 \\ 
 3000 & 30 & 1.0 & 14 \\ 
  & & 2.0 & 25   \\
    & & 3.0 & 31 \\ \hline
   & & 0.5 & 10 \\ 
   3000 & 10 & 1.0 & 25  \\
 & & 2.0 & 31\\
    & & 3.0 & 43\\ \hline
   & & 0.5 & 26 \\ 
  1000 & 30 & 1.0 & 33 \\
  & & 2.0 & 36 \\
    & & 3.0 & 38\\ \hline
   & & 0.5 & 80 \\ 
  1000 & 10 & 1.0 & 85 \\
 &  & 2.0 & 89  \\
    & & 3.0 & 92 \\ \hline
 
\end{tabular}
\end{table}

\subsection{Application to Real Data}\label{sec:data}

We now investigate the effects of observation on the intrinsic emission line properties of real \lya sources. We use archival data to explore recovery in high and low resolution data sets. Here we focus on the recovery of skew so as to explore its utility as an indicator of galaxy dynamics. As we have seen, FWHM is fully recoverable in the entirety of the tested parameter space given a correction for systematic offsets, and its recovery is thus explored no further.

\subsubsection{FORS2 - Low Resolution}\label{sec:fors2}

To explore properties of a large galaxy sample, low resolution catalogues may be a valuable source. Here we sample the VLT/FORS2 GOODS catalogue data of resolution $\mathrm{R = 660}$  \citep[][]{vanzella05,vanzella06}. This catalogue includes 1165 spectroscopically confirmed redshifts, with 131 confirmed \lya emission line sources at $\mathrm{ 2.5\,<\,z\,<\,6.0}$. However, as seen in section \ref{sec:results}, lower resolution detection can lead to poor characterisation of skew. By fitting the confirmed \lya features in the archival 1D extraction spectra we produce a distribution of skew measurements for the sample. We  match the measurements to new $\mathrm{R = 660}$ recovery plots and find the sub-sample of sources for which intrinsic skews is unambiguously recoverable. We show this in figure \ref{fig:fors2samp}. We exclude any measurements within the instrumental resolution width of a sky line, 17\% of the sample. 

\begin{figure}
\includegraphics[scale=0.33]{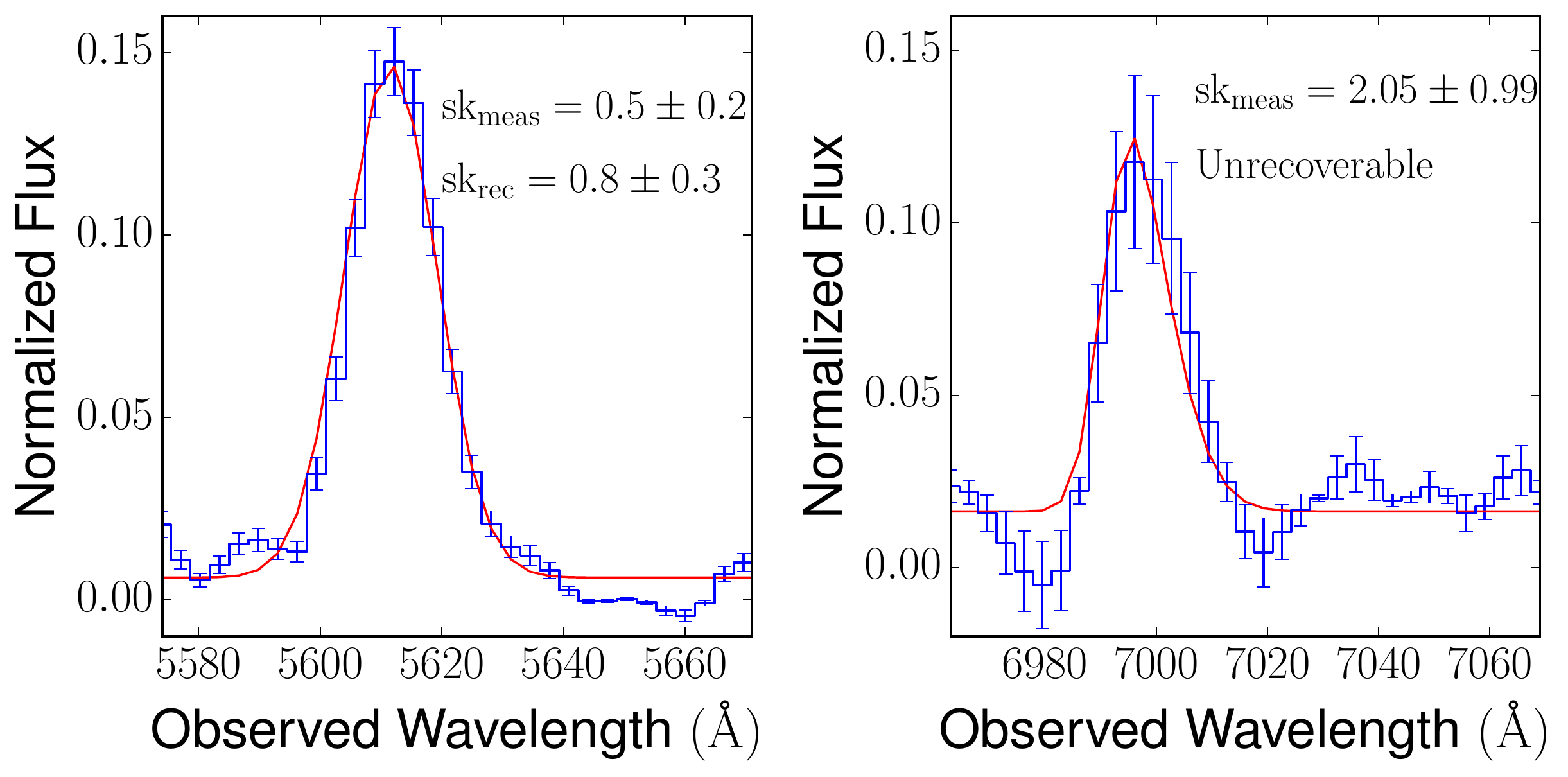}
\caption{Examples of two GOODS \lya observations taken with the VLT/FORS2 instrument. On the left is an example of a  line with recoverable skew, on the right a line with unrecoverable parameters is shown. Measured skew and, where applicable, recovered intrinsic skew are labeled on each plot. In this example, it is not low signal-to-noise, but rather high measured skew and narrow line velocity width which render the skew parameter unrecoverable.}
\label{fig:skew_ex}
\end{figure}

From the sample of \lya emission sources we determine that only 13\% have recoverable intrinsic skews, based on the model calculations shown in figures \ref{fig:sktrk100}-\ref{fig:veltrk10}, with an average uncertainty of $\mathrm{\sigma_{int} = 42\%}$. We illustrate an example of recoverable and unrecoverable emission lines  in figure \ref{fig:skew_ex}. The recoverable lines tend to have high signal-to-noise and relatively low skew values. A line may have an unrecoverable skew for one of several reasons:  there may be multiple components to the emission line which require a very skewed measured Gaussian, the line may be of low signal to noise, or it may merely have a sufficiently high intrinsic skew or a sufficiently low velocity width that the deconvolution with the spectral resolution leads to degeneracy. The example in figure \ref{fig:skew_ex} falls in the last of these categories, with the well-fit best model being both narrow and skewed - a difficult situation in which to recover the intrinsic properties. These values are heavily biased towards high velocity width and S/N emission profiles and only the low intrinsic skew distribution can be successfully probed. As a result we see the tail of large skew values measured in the original distribution is not reliably characteristic of the galaxy properties, and is caused by measurement noise and proximity to sky emission line residuals. This result precludes the use of low resolution datasets in probing well-sampled distributions of intrinsic skew.

\subsubsection{DEIMOS - High Resolution}\label{sec:deimos}

To explore high resolution and high S/N data, an archival sample of KECK/DEIMOS \citep[][]{faber03} observational data was taken from \cite{mallery12}. The data set was observed in the $\mathrm{600\,line\,mm^{-1}}$ grating mode. This gives an instrumental resolution of $\mathrm{1.4\,\angstrom}$ ($\mathrm{R \sim 3000}$). The data set includes 244 \lya sources at 4.2\,$<\,z\,<$\,5.6. In addition to FWHM and skew information, \citet{mallery12} report star formation rate (SFR), stellar mass and stellar population age calculated through fitting the galaxy spectral energy distributions (SED), using \cite{bruzualcharlot03} spectral synthesis models.

We again match the measured data with recovery plots and find the inferred intrinsic skews, and filter for possible sky line masking contamination (7\% of the sample). We show these results in figures \ref{fig:deimdata} and \ref{fig:deimcdf}. From the sample of 244 \lya sources, 147 $\mathrm{(\sim 60\%)}$ had recoverable skews, with an average uncertainty of $\mathrm{\sigma_{int} = 27\%}$. As was the case for the low resolution sample, we see a reduction in the large skew tail and a slight bias towards higher velocity widths and S/N. The original measured sample contained 8 sources with high skew values, which we define as $\mathrm(sk > 5.0)$. These are unphysical given this instrumental resolution. Only one was within the instrumental resolution width of a sky line. This suggests that high skew tails can be generated purely through measurement noise. The median skews of the measured and recovered intrinsic sample are $1.41\pm0.39$ and $1.31\pm0.76$ respectively. The corresponding mean skews are $1.44\pm0.31$ and $1.32\pm0.61$. We see that the initial distribution is broadly preserved, although the number of sources with reliably recovered intrinsic skews is lower than that with measured values.

\subsection{Physical Interpretation}

\begin{figure*}
  \centering
  \begin{minipage}[t]{0.47\textwidth}  
	\includegraphics[width=\textwidth]{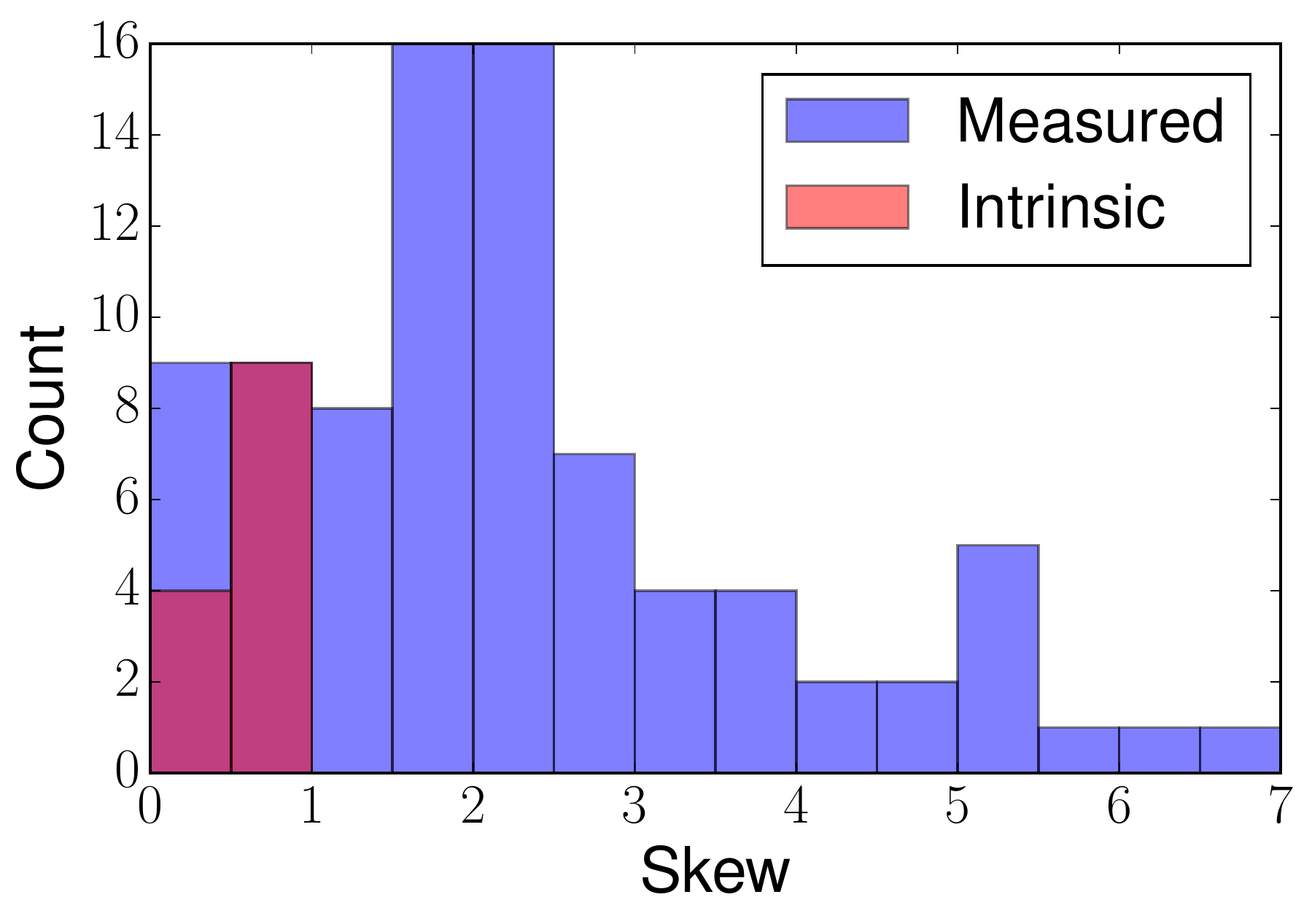}
	\caption{\lya emission lines at $R=660$ drawn from the  VLT/FORS2 archive (ESO GOODS, PI:Vanzella) showing the line skew parameter distribution. Shaded in blue are measured profile skews, in red are the recovered intrinsic skews.}
	\label{fig:fors2samp}  
	\end{minipage}
    \hfill
  \begin{minipage}[t]{0.49\textwidth}
    \includegraphics[width=\textwidth]{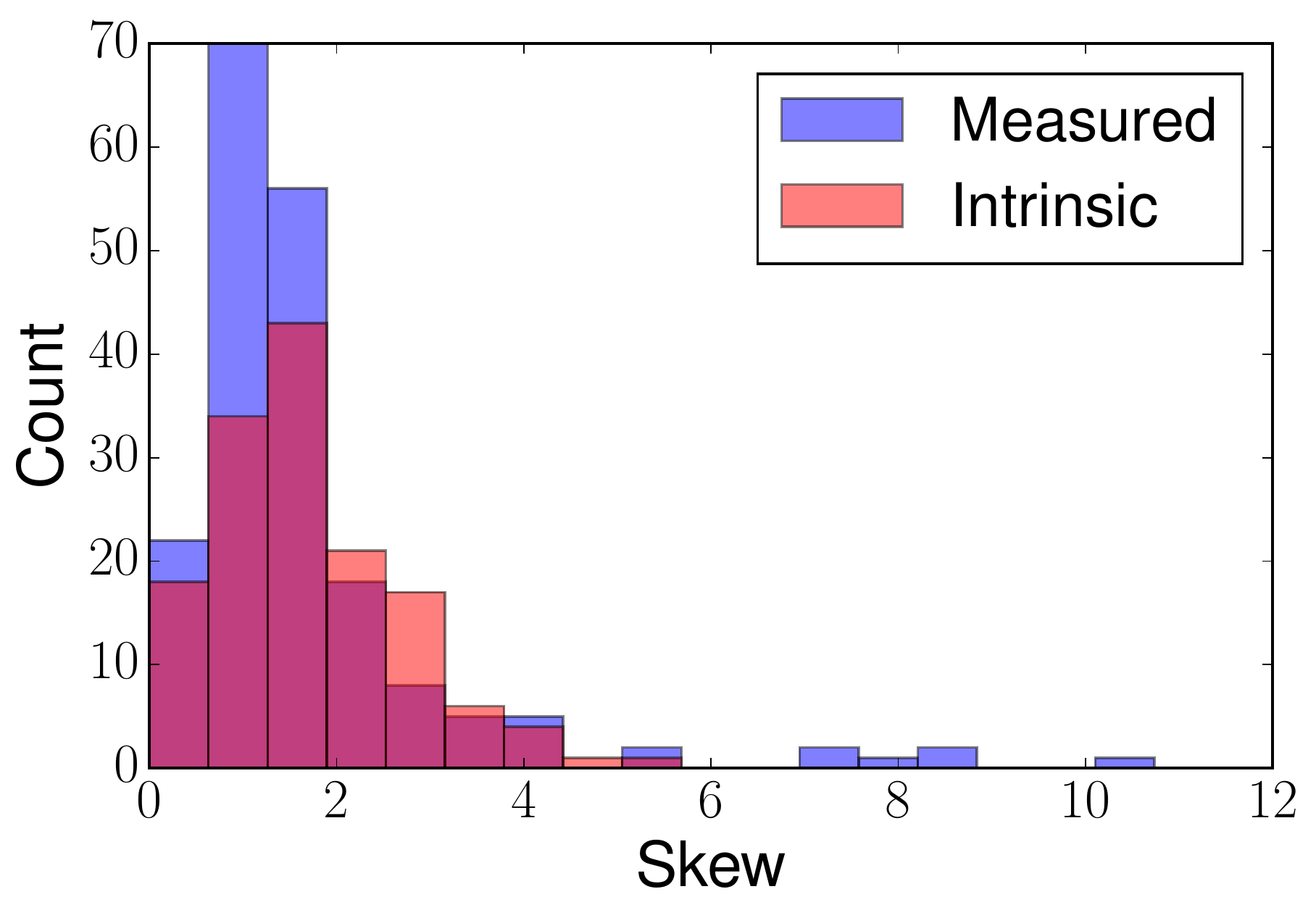}
    \caption{\citet{mallery12}  \lya emission line sample skew distribution as measured with Keck/DEIMOS at $R\sim3000$. Shaded in blue are measured profile skews, in red are the recovered intrinsic skews.}
    \label{fig:deimdata}
  \end{minipage}

  \centering
  \begin{minipage}[c]{0.47\textwidth}
    \includegraphics[width=\textwidth]{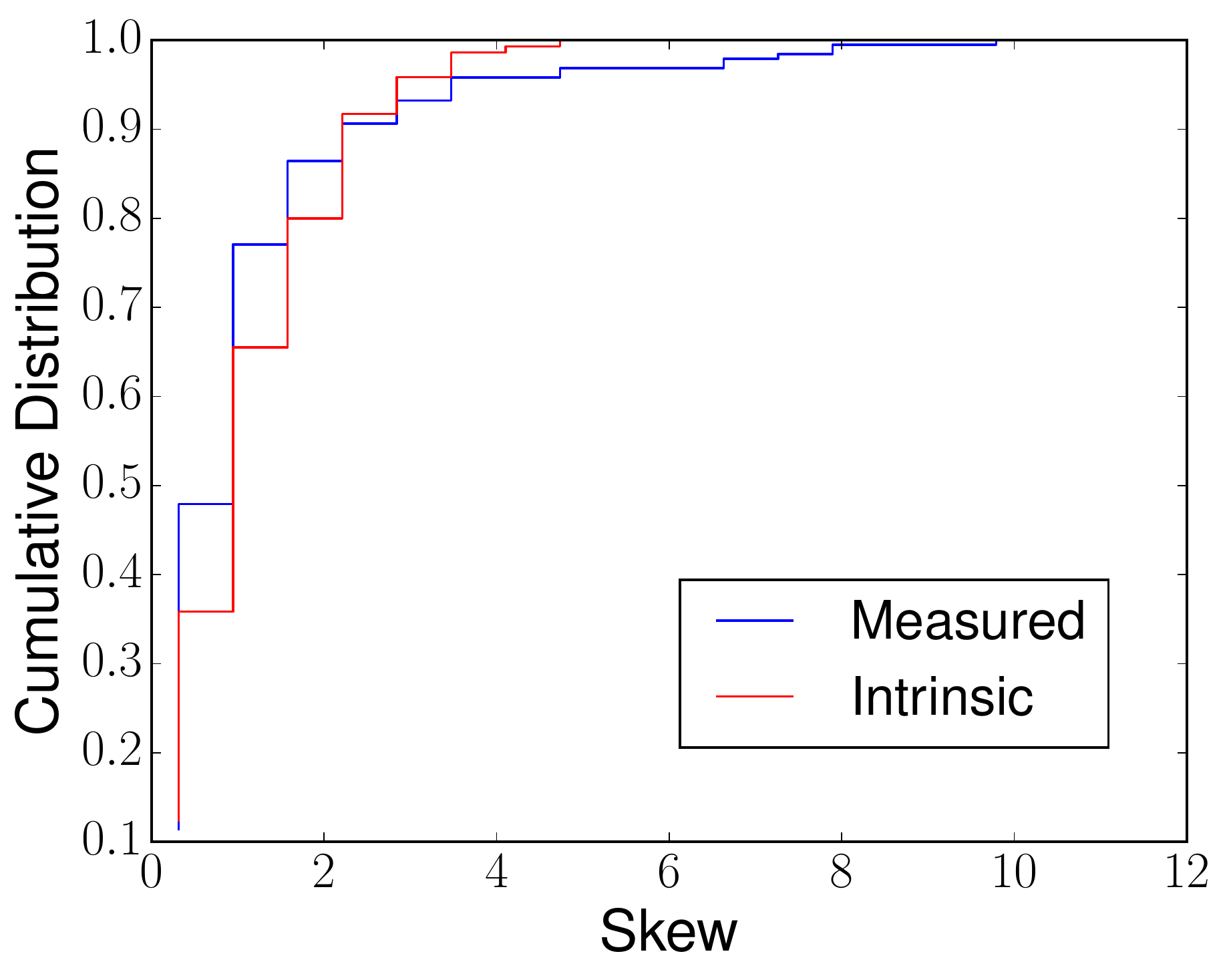}
    \caption{Cumulative distribution of skews in the \citet{mallery12}  DEIMOS \lya emission sample. In blue is the measured profile distribution, in red is the recovered intrinsic profile distribution.}
    \label{fig:deimcdf}  
  \end{minipage}
  \hfill
  \begin{minipage}[c]{0.47\textwidth}
    \includegraphics[width=\textwidth]{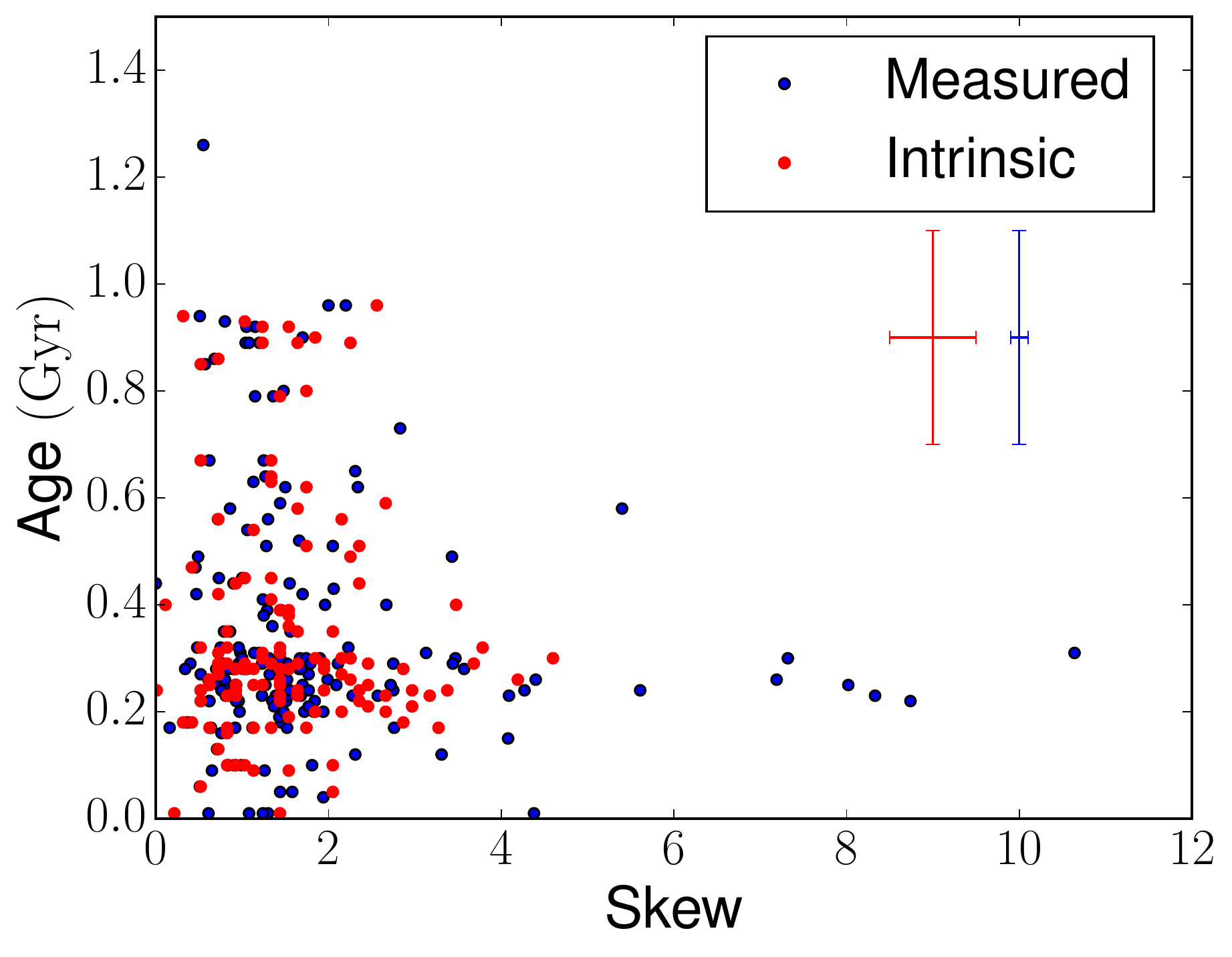}
    \caption{Distribution of the \citet{mallery12}  DEIMOS sample in stellar population age in Gyr vs. skew. Points indicate measured (blue) and intrinsic (red) skew values. Representative example error bars are shown as individual points.}
    \label{fig:m12age}
  \end{minipage}

  \centering
  \begin{minipage}[b]{0.47\textwidth}
    \includegraphics[width=\textwidth]{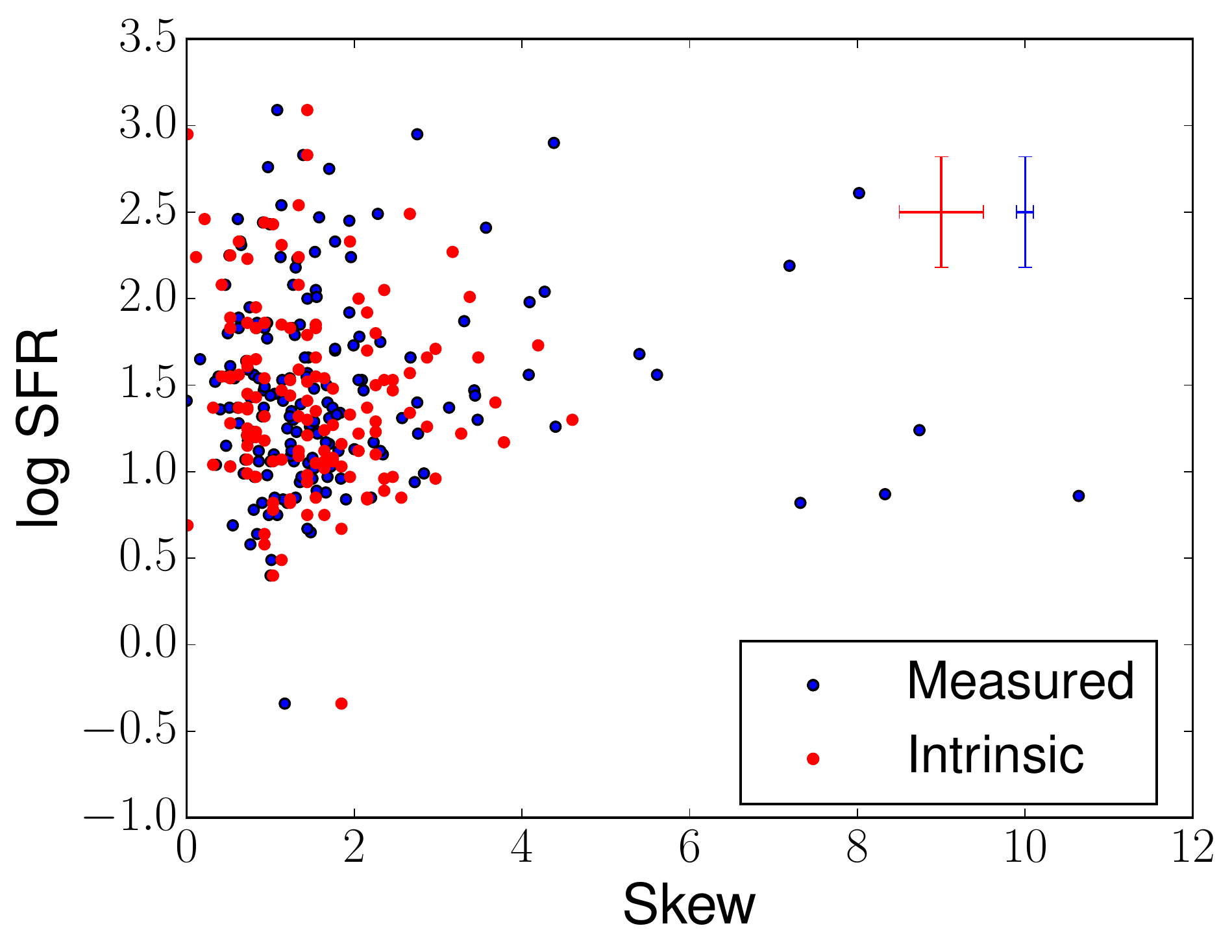}
    \caption{\citet{mallery12}  DEIMOS sample distribution in log(SFR in M$_\odot$\,yr$^{-1}$) vs. skew. In blue is the measured profile distribution, in red is the recovered intrinsic profile distribution.}
    \label{fig:m12lsfr}
  \end{minipage}
\hfill
\begin{minipage}[b]{0.47\textwidth}
    \includegraphics[width=\textwidth]{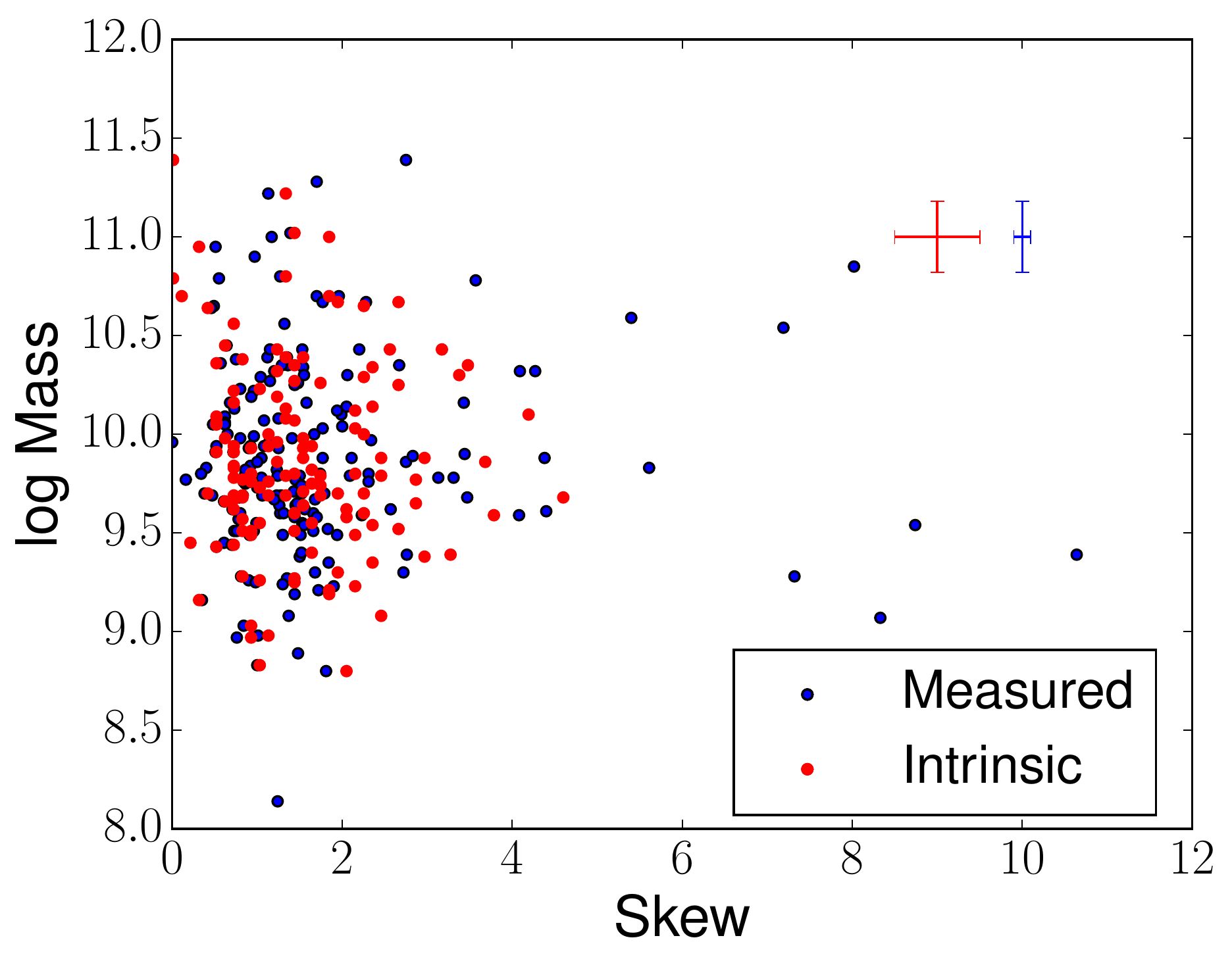}
    \caption{\citet{mallery12}  DEIMOS sample showing log(stellar mass in M$_\odot$) vs. skew. In blue is the measured profile distribution, in red is the recovered intrinsic profile distribution.}
    \label{fig:m12lm}
  \end{minipage}
\end{figure*}

\begin{figure*}
  \centering
  \begin{minipage}[t]{0.8\textwidth}  
	\includegraphics[width=\textwidth]{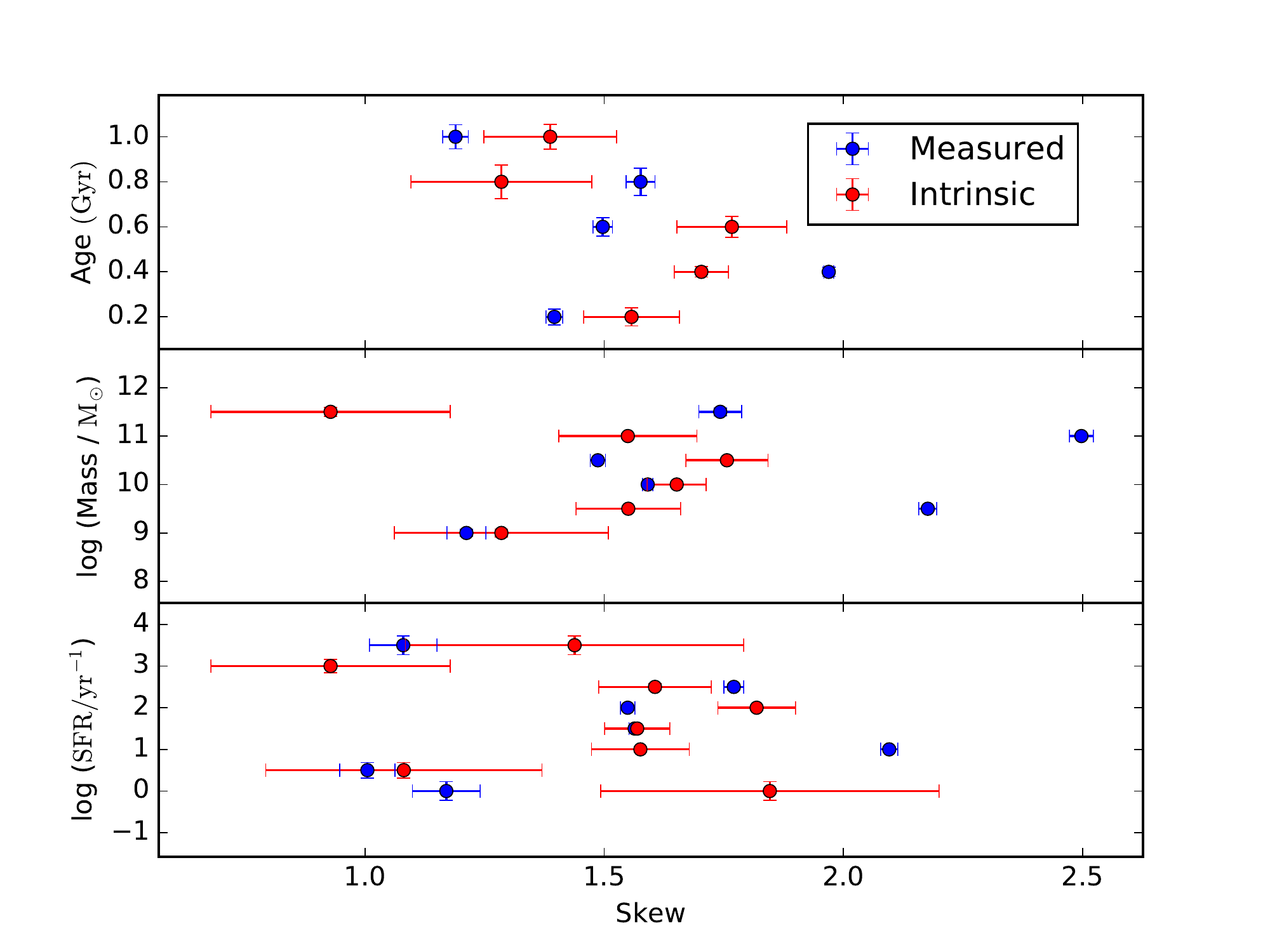}
	\caption{\citet{mallery12}  DEIMOS sample binned by SED-derived physical characteristics to explore any dependence on skew. In blue is the measured profile distribution, in red is the recovered intrinsic profile distribution.}
	\label{fig:m12sks}  
	\end{minipage}
    \hfill
  \begin{minipage}[b]{0.8\textwidth}
    \includegraphics[width=\textwidth]{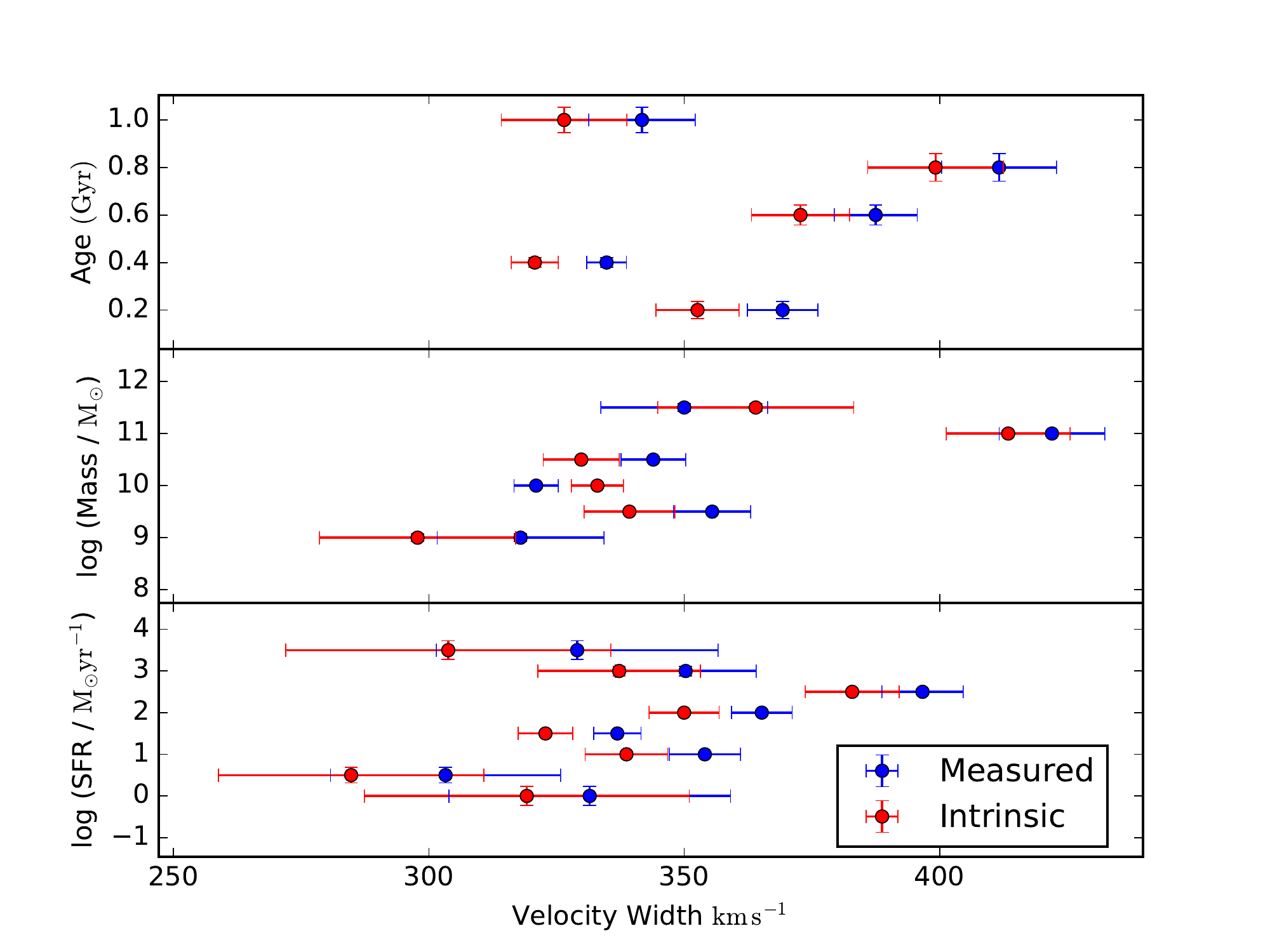}
    \caption{\citet{mallery12}  DEIMOS sample binned by physical characteristics to explore any dependence on velocity width. In blue is the measured profile distribution, in red is the recovered intrinsic profile distribution.}
    \label{fig:m12fws}
  \end{minipage}
\end{figure*}

If we expect skew to be a possible indicator of outflows we may expect it to correlate with other galaxy properties such as age, mass and star formation rate \citep[e.g.][]{steidel10,yamada12,trainor15}. The asymmetry of the \lya emission line is generated by the diffusion of line photons through a galaxy's warm ISM and circumgalactic material. The more scattering events the \lya photons undergo, the more the asymmetric wings are broadened. In galaxies with outflows this effect will be most pronounced when looking down the line of sight of an outflow. The asymmetry might be expected to increase with SFR, as a larger young stellar population will drive stronger outflows, although an increase in mass and hence gravitational potential will diminish the effect. We may also expect a correlation with age, as an older star-forming galaxy will have ionised a larger region of its surroundings. However, this simplistic overview ignores the effect of more complex galaxy structure.\citet[][]{steidel10} did not find any significant correlation between outflow velocities and the galaxy star formation rate, whereas \citet[][]{trainor15} found a weak dependence between the two. The discrepancy between these studies is explained by the difference in SFR regimes spanned by the samples probed, so care needs to be taken to understand the context in which any given sample is placed. There are likely to be effects on both the star formation and any escaping \lya emission due to ongoing versus bursty star formation, or due to complexities in the merger history of a given galaxy. Patchy star-formation, for example, may result in regions with enhanced outflows which may, in turn, present larger \lya skews, whereas a disrupted or complex gas morphology may suppress the bulk motions assumed in the simple model of spherical gas outflows discussed above.

The combination of all the effects discussed above means that any correlation between skew and outflows will be difficult to interpret in the absence of additional morphological or star formation history information.  Any intrinsic relation between two parameters, such as star formation rate or age and \lya line shape, will be masked by the additional effects of  star formation history, gas composition and dynamics and other physical parameters. The resonant scattering of \lya renders it particularly vulnerable to these effects, in addition to any effects of instrumentation. Nonetheless, the possibility of a trend between \lya line shape and galaxy properties remains. Thus it would be useful to evaluate the presence or absence of any such relation between spectral parameters and galaxy properties in a suitable observed data set, and interpret such with care.

Using the \citet{mallery12} sample we can compare the measured and intrinsic skew values derived here with the galaxy properties also calculated by ~\cite{mallery12}. We present these comparisons in figures \ref{fig:m12age}-\ref{fig:m12lsfr}. When plotting SED-inferred age of each galaxy against skew, the measured values suggest a weak evolutionary trend such that scatter in skew decreases at larger ages. This could be produced as older stellar populations will drive out a large ionisation bubble thereby reducing the \lya back-scatterings. We divide the data, shown in figure \ref{fig:m12age}, into two bins from age $\mathrm{0-0.5 \,Gyr}$ and $\mathrm{>0.5 \,Gyr}$, with sample sizes of 185 and 43 respectively. A KS test of the two sub-samples gives a 67\% probability of the null hypothesis, that the two samples are drawn from the same distribution. Thus any possible trend is very weak, and reliant on the high skew wings which have been shown at this resolution to be almost entirely generated by noise. When the recovered intrinsic skew distribution is considered, the trend completely disappears as the tails of the skew distribution are removed. There is also an increase in average uncertainty on each point. While no trend is seen up to the maximum skew that can be recovered, we note that very high intrinsic skews would be impossible to recover at this instrumental resolution. Thus we cannot preclude the possibility of a high skew tail at young ages, but neither does the data show any robust evidence for one. We see no correlation when SFR is plotted against skew in figure \ref{fig:m12lsfr}. Similarly there is no correlation between mass and skew, as shown in figure \ref{fig:m12lm}. 

We also bin the data by its physical characteristics and compare with the mean skew or FWHM in each bin. We see no bulk correlations between the sample galaxy properties with skew, see figure \ref{fig:m12sks}. When we plot the binned data with velocity width, in figure \ref{fig:m12fws}, we see a weak apparent trend with mass. 

\subsection{Blue-Red flux asymmetry}

As discussed above, the unreliable characterisation of the skew parameter defined in equation (\ref{eq:skgauss}), at common instrumental resolutions limits its utility as an indicator of galaxy dynamics. We now investigate whether a simpler parametrisation could be used as a more reliably recoverable measurement of asymmetry. For this we use the ratio of the total blue and red emission wing fluxes, measured either side of the peak of the emission line, which we denote B/R.

By taking our simulated  models of skewed \lya lines we can calculate both the recovered and measured B/R values corresponding to a known input skew. As the central peak returned by the skewed Gaussian fit is not entirely reliable, especially at low S/N, we use the maximum of each spectrum within $\mathrm{1 \sigma}$ of the modelled central wavelength as the peak of the emission line. We then measure the integrated flux either side of this peak wavelength out to $\mathrm{\pm 1000 \,km\,s^{-1}}$ to obtain the B/R flux ratio. While our models are set at a flat zero continuum, meaning wavelengths outside the emission line will average to zero flux, the inclusion of wavelengths beyond those occupied by the line will increase measurement noise. By performing this method across our previous parameter space, we can observe how B/R is affected by modelled instrumentation. 

\begin{figure}
  \centering  
	\includegraphics[width=0.50\textwidth]{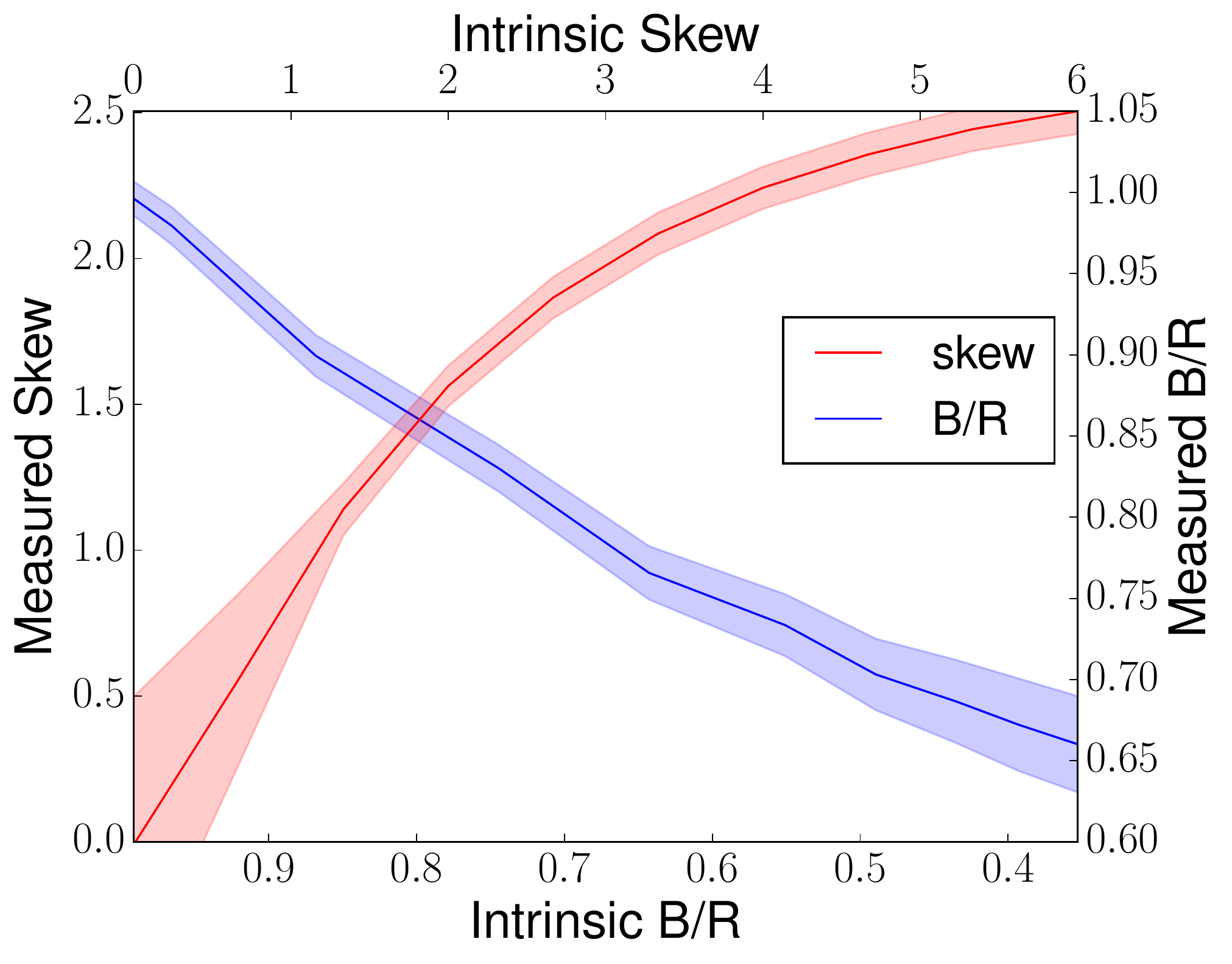}
	\caption{Comparison of measured and intrinsic B/R and skew asymmetry parameters, for an emission line with  $\mathrm{FWHM\,=\,250\,km\,s^{-1}}$, measured with R = 3000 and S/N = 30.}
	\label{fig:BRcomp}  
\end{figure}

As an example we consider a single line at $\mathrm{FWHM\,=\,250\,km\,s^{-1}}$, and plot how the difference between intrinsic and measured values in both skew and B/R evolves for a simulated detection at $\mathrm{R = 3000}$ and $\mathrm{S/N = 30}$ in figure \ref{fig:BRcomp}. While the intrinsic skew is associated with a large uncertainty for measured skews $<1$, the B/R flux ratio remains tightly constrained, with a near-linear relation between intrinsic and measured B/R value. For highly skewed lines, the measured skew becomes largely insensitive to the line profile. In this regime B/R remains well constrained, and the intrinsic value can be reconstructed from the measured value, but measurements will systematically underestimate the line asymmetry unless adjusted for instrumental effects.

As expected, the simple parametrisation shows a larger overall recoverability, in addition to being entirely non-degenerate in the example shown. As a result, while skew may show correlations with galaxy properties, its difficulty in reconstruction can make it challenging to interpret, whereas B/R remains a well-characterised parameter. This highlights the fact that while more complex parametrisations may be possible, observational constraints may mean that these over-interpret the true information content of the spectra. In this case a simple flux ratio may be more robust. Unfortunately we are unable to explore the relation between physical properties and B/R directly since those datasets for which the former is available do not provide the latter, and vice versa. This would be a fruitful avenue for future studies in the increasing archive of large, uniform galaxy spectroscopic samples which are now starting to publicly release data, but any quantitative analysis will require very large samples, good signal-to-noise, robust and uniform photometry and good spectral resolution - a combination which is challenging to achieve. 

\subsection{Future Instrumentation}

We have seen thus far that with modern instrumentation, the accurate recovery of \lya spectral line profiles is possible but limited, most notably in skew. We now explore the effect of future instrumentation on the recovery of these parameters. Future projects of note that will allow unprecedented \lya mapping of the night sky include NIRSPEC on the {\em James Webb Space Telescope} (JWST) and EAGLE on the European Extremely Large Telescope (E-ELT).

The NIRSPEC instrument on JWST is an optical-NIR spectrometer that will span a wavelength range of $\mathrm{0.6-5\mu m}$. This would allow for \lya detection at redshifts in excess of $\mathrm{z\sim10}$. NIRSPEC will have a high resolution mode of $\mathrm{R=2700}$ \citep{birkmann16}. At a central \lya emission wavelength $\mathrm{\lambda\sim1.0\mu m}$ (observed frame) this corresponds to a \lya redshift of $\mathrm{z\sim7.2}$ and a wavelength resolution of $\mathrm{\Delta\lambda=3.74\,\angstrom}$, a limiting velocity width of $\mathrm{110\,km\,s^{-1}}$. This suggests NIRSPEC will have an adequate resolution to successfully characterize \lya skew in most sources, although for particularly narrow lines and at very high redshifts S/N will be a significant limiting factor.

The EAGLE instrument for E-ELT is a NIR spectrometer with a proposed maximum resolution mode of $\mathrm{R\sim10000}$ \citep{morris12}. In a similar case to that explored with NIRSPEC, a line at $\mathrm{1\,\mu m}$ would be measured with a limiting instrumental velocity width of $\mathrm{30\,km\,s^{-1}}$. This means that higher redshift \lya skew may be reliably recovered with EAGLE even in relatively small galaxies.  

\begin{figure}
  \centering  
	\includegraphics[width=0.48\textwidth]{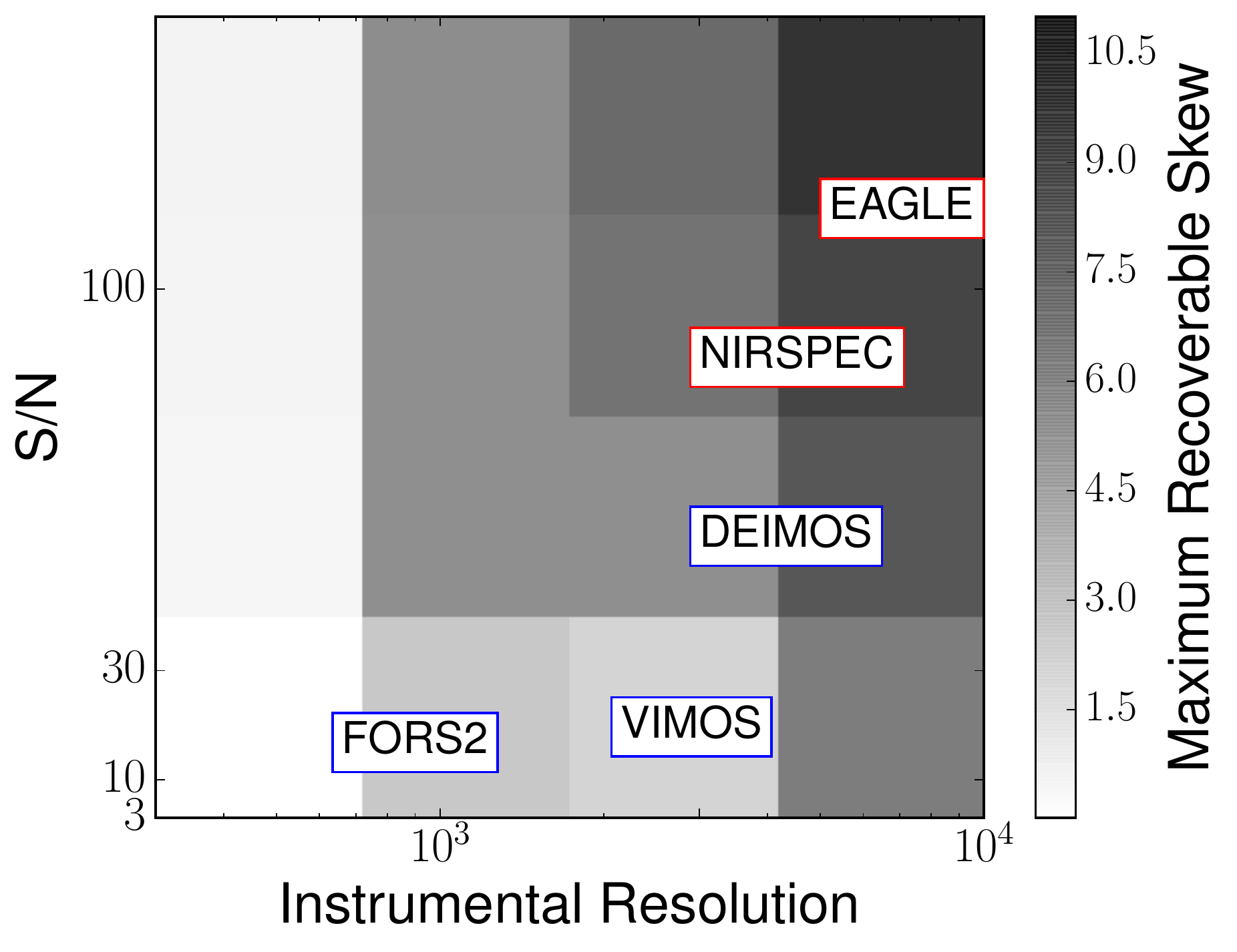}
	\caption{Maximum reliably recovered intrinsic skew in S/N and resolution parameter space, for an emission line with  $\mathrm{FWHM\,=\,250\,km\,s^{-1}}$. Example 3.5 hour exposures of a $\mathrm{m_{AB}=24}$ LAE at either $z=3$ (optical) or $z=7$ (infrared) are shown for selected existing and forthcoming instruments.} Observations in the visible spectrum are outlined in blue, whereas observations in the infrared are outlined in red.
	\label{fig:instrs}  
\end{figure}

We conclude this analysis by plotting all instruments discussed in this work on a contour plot of maximum recoverable skew. By considering an example galaxy with a \lya emission line of $\mathrm{FWHM\,=\,250\,km\,s^{-1}}$, and $\mathrm{m_{AB} = 24.0}$, we make use of the relevant instrumental exposure time or integration time calculators to explore the relative strengths of different instruments. In figure \ref{fig:instrs} we demonstrate how the maximum recoverable skew limit varies, based on the instrument used and with a uniform exposure time of 3.5 hours. We see, as expected, that the high resolution modes of FORS2 and VIMOS both produce S/N values too low for accurate skew recovery. The high resolution modes of DEIMOS, NIRSPEC and EAGLE produce maximum recoverable skews of $\mathrm{sk\,>\,4.5}$. As seen in section \ref{sec:deimos}, this allows for reliable recovery of intrinsic line asymmetry and FWHM distributions. 

\section{Conclusion}\label{sec:conclusions}

In this paper we explore the transformation of intrinsic \lya spectral properties through simulated observations.
By convolving a set of generated \lya emission lines of known line profile with different instrumental resolutions and adding a variety of noise sources, the effect of detection limitations on measured properties was explored.
The effect of sky line masking was also discussed, as well as how the results affect the interpretation of both real and possible future data sets. Our main results are summarized as follows:

\begin{enumerate}

\item While \lya asymmetry, characterised through a skewness parameter, could be measured for any resolution, at resolutions below $R=3000$ and velocity widths of $250$\,km\,s$^{-1}$ and below, the implied intrinsic skew becomes degenerate and un-recoverable. This was seen both in simulated and in real data, where a fraction of sources had unrecoverable intrinsic skews.

\item The inclusion of sky line masking introduces a deviation in the measured skew of lines overlapping a masked region, with offsets between measured and intrinsic skew reaching 40\% for a sky line-affected example at $R=3000$.

\item When applied to real data sets, the relatively low resolution FORS2 instrument was unable to characterise the intrinsic skew distribution, with only 13\% of the values being recoverable in a measured sample. A higher resolution DEIMOS sample showed 60\% recoverability, as well as being able to probe the distribution successfully.

\item We observed how skew trends with galaxy properties in the DEIMOS sample. Mass, SFR and age show no statistically reliable trends, in measured or intrinsic skew. This suggests that skew is not a reliable independent indicator of the presence of outflows or other galaxy properties.

\item When compared to simple blue-to-red flux ratio parametrisation of asymmetry we see, as expected, that B/R produces a better constraint on intrinsic asymmetry, as it does not become degenerate in the sample space.

\item Future instrumentation will push the detection of \lya to ever higher redshifts. Considering the instrumental resolutions that will be achieved in JWST/NIRSPEC and E-ELT/EAGLE we find that the recovery of intrinsic skew values will be possible for a much broader range of galaxies.

\end{enumerate}

We note that the data set used for evaluation purposes in this paper did not include sky-projected morphological information on each galaxy. In the future, with the increasing availability of IFU spectroscopy, including IFU instrumentation with adaptive optics capability, it may be possible to relate Lyman-$\alpha$ spectral properties to galaxy morphology not only in two dimensions, but in three.  While this may help clarify the physical interactions between gas and light which shape this highly complex emission line, the analysis carried out in this paper makes it clear that instrumental limitations and good signal-to-noise will be key in forming a reliable, detailed picture of such interactions.

\section{Acknowledgements}

We acknowledge funding from UK Science and Technology Facilities Council (STFC) consolidated grant ST/L000733/1 and PhD studentship 1622079. This work includes analysis of data from obtained from the ESO Science Archive Facility. We made use of Ned Wright's Javascript Cosmology Calculator \citep{2006PASP..118.1711W} and also the very useful TopCat catalog manipulation tool \citep{2005ASPC..347...29T}. This research made use of Astropy, a community-developed core Python package for Astronomy \citep{2013A&A...558A..33A}.


\bsp	
\label{lastpage}
\end{document}